\documentclass[twocolumn,trackchanges]{aastex7}
\usepackage{lmodern}
\usepackage{ae,aecompl,multirow}
\usepackage{threeparttable}
\usepackage{booktabs}
\usepackage{url}
\usepackage{amsmath}
\usepackage{graphics,comment}
\usepackage{textcomp} 

\usepackage{hyperref}

\usepackage{booktabs}  
\DeclareUnicodeCharacter{2212}{\textminus}
\begin{document}

\title{Decomposition of disk and jet optical emission: Case of PKS~1510-089 and PKS~0736+017}

\correspondingauthor{Shivangi Pandey}
\author[0000-0002-4684-3404]{Shivangi Pandey}
\affiliation{Center for Theoretical Physics, Polish Academy of Sciences,
Al. Lotników 32/46, 02-666 Warsaw, Poland}
\email[show]{spandey@cft.edu.pl}
\affiliation{Aryabhatta Research Institute of Observational Sciences, Nainital\textendash263001, Uttarakhand, India}
\affiliation{Department of Applied Physics/Physics, Mahatma Jyotiba Phule Rohilkhand University, Bareilly\textendash243006, India}

\author[0000-0002-8377-9667]{Suvendu Rakshit}
\affiliation{Aryabhatta Research Institute of Observational Sciences, Nainital\textendash263001, Uttarakhand, India}
\email{suvenduat@gmail.com}

\author[0000-0003-2558-1406]{Adarsh Kumar}
\affiliation{Institute of Astronomy and Planetary Sciences, Universidad De Atacama, Av. Copayapu 485, 1531772 Copiapó, Atacama, Chile}
\email{adarsh.kumar.25@alumnos.uda.cl}

\author[0000-0002-0393-0647]{Anne L\"ahteenm\"aki}
\affiliation{Aalto University Mets\"ahovi Radio Observatory, Mets\"ahovintie 114,
02540 Kylm\"al\"a, Finland}
\affiliation{Aalto University Department of Electronics and Nanoengineering, 
P.O. BOX 15500, FI-00076 AALTO, Finland.}
\email{anne.lahteenmaki@aalto.fi}

\author[0000-0003-1249-6026]{Merja Tornikoski}
\affiliation{Aalto University Mets\"ahovi Radio Observatory, Mets\"ahovintie 114,
02540 Kylm\"al\"a, Finland}
\email{merja.tornikoski@aalto.fi}

\author{Archana Gupta}
\affiliation{Department of Applied Physics/Physics, Mahatma Jyotiba Phule Rohilkhand University, Bareilly\textendash243006, India}
\email{archana.gupta@mjpru.ac.in}

\begin{abstract}
Measuring BLR size and black hole mass of jetted quasars with strong radio emission is challenging due to the contribution from non-thermal synchrotron radiation from the jet in the optical continuum. Here, we investigate the optical continuum emission in two flat-spectrum radio quasars (FSRQs), PKS~0736+017 and PKS~1510-089, by separating the thermal emission from the accretion disk and the non-thermal synchrotron emission from the jet using their distinct polarization signatures. Our analysis combines multi-epoch optical spectro-photometric and polarimetric observations from the archival optical continuum, H$\beta$ emission line light curves from earlier reverberation mapping campaigns, and 37 GHz radio data. For PKS~0736+017, we find that the disk contributes approximately 70\% ($f_d = 0.70^{+0.01}_{-0.01}$) of the total optical emission. After removing the jet component, we obtain a revised H$\beta$ lag of $84.8^{+23.2}_{-18.4}$ days. In contrast, the disk contribution in PKS~1510-089 is lower with 37.5\% ($f_d = 0.37^{+0.01}_{-0.01}$), with a corresponding H$\beta$ lag of $54.4^{+4.4}_{-5.4}$ days. Our findings highlight the need for careful decomposition of optical light curves in jet-dominated AGN.
\end{abstract}
\section{Introduction}
Flat-spectrum radio quasars (FSRQs) are a subclass of blazars characterized by their relativistic jets oriented close to our line of sight, leading to significant Doppler boosting of the jet emission. This orientation not only amplifies non-thermal continuum variability but also often obscures the intrinsic thermal emission from the accretion disk, complicating interpretations of its optical and ultraviolet properties. In the context of Reverberation Mapping (RM) \citep{Bahcall1972, Blandford1982, Peterson_1993}, which is a traditional method for studying the central engine of AGNs, uses the time delay between variations in the ionizing continuum (from the accretion disk) and the broad emission lines (from the BLR) to measure the size of the BLR, this jet dominance introduces major challenges.

A recent investigation by \citet[][hereafter Li2020]{Li2020} on the FSRQ 3C~273 \citet{Hazard2018} demonstrated that jet contamination can substantially bias RM results. Through a long-term spectroscopic monitoring campaign, they found that the optical continuum of 3C~273 includes a slowly varying component that does not correlate with broad-line variations. By decomposing the light curves into jet and disk components, they showed that up to 28\% of the optical flux could originate from the jet, thereby diluting the reverberation signal. Their work highlighted the importance of isolating the thermal continuum when applying RM to jet-dominated AGNs and set a precedent for such decomposition in other FSRQs.

Motivated by these findings, we investigate two additional $\gamma$-ray bright FSRQs, PKS~0736+017 and PKS~1510-089, both of which are known for strong multiwavelength variability and pronounced jet activity. PKS1510, a powerful FSRQ at $z = 0.361$, shows extreme flux variability across the entire electromagnetic spectrum. In an 8.5-year RM campaign, \citet[][hereafter R2020]{Rakshit2020} measured a rest-frame H$\beta$ lag of $\sim$45 days, and a black hole mass of $5.71^{+0.62}_{-0.58} \times 10^7 M_\odot$. Multiple prominent optical flares coinciding with gamma-ray activity provide clear evidence of significant jet contamination in the optical continuum, likely due to synchrotron emission from the jet. This non-thermal contribution complicates the use of the optical continuum as a reliable tracer of the ionizing luminosity. It necessitates correction when estimating the size of the broad-line region (BLR). 

Similarly, PKS0736, at redshift $z = 0.189$, displays a double-peaked spectral energy distribution (SED) typical of blazars and exhibits significant non-thermal contamination in the optical band. \citet[][hereafter P2022]{Pandey2022} conducted a spectroscopic RM campaign of this source and reported a rest-frame H$\beta$ lag of $\sim$66 light-days, leading to a black hole mass estimate of $7.3 \times 10^7 M_\odot$. The presence of strongly correlated flares in both the optical and gamma-ray bands indicates a substantial non-thermal contribution from the jet, particularly during high-activity states. Despite this, the emission-line light curves, especially H$\beta$, show a measurable lag relative to the optical continuum. Notably, both sources deviate from the canonical $R_{\mathrm{BLR}}$–$L_{5100}$ relation, reinforcing the need to account for non-thermal contamination when interpreting their RM results.

In this study, we revisit and expand upon these previous analyses by disentangling the thermal and non-thermal contributions to the optical continuum in both PKS0736 and PKS1510. By isolating the thermal disk component, we aim to refine the BLR size estimates and assess the extent to which jet variability affects the applicability of RM-based black hole mass measurements in FSRQs. To achieve this, we employ \texttt{DECOMP}, a Bayesian decomposition framework developed by Li2020, which models the total observed optical emission as a superposition of two independent stochastic processes representing the disk and jet contributions. By jointly fitting the optical continuum, broad emission-line light curves, radio data, and polarization measurements, \texttt{DECOMP} effectively separates the disk and jet components, providing robust estimates of their temporal and physical properties.

The structure of this paper is as follows. In Section~\ref{sec:data}, we describe the multi-wavelength monitoring data collected for both sources. Section~\ref{sec:Jet evidence} outlines the motivation for investigating jet dominance in the optical continuum emission of these objects. The methodology and operational framework of the \texttt{DECOMP} code are presented in Section~\ref{sec:DECOMP}. The results of the analysis are detailed in Section~\ref{sec:results}, followed by a discussion in Section~\ref{sec:discussion} and concluding remarks in Section~\ref{sec:conclusion}.
\section{Data}
\label{sec:data}

\begin{table*}
\caption{Summary of Observations for PKS~0736+017 and PKS~1510-089}
\centering
\begin{tabular}{lcccccc}
\hline
\multirow{2}{*}{\textbf{Source}} & \multirow{2}{*}{\textbf{Telescope/Instrument}} & \multirow{2}{*}{\textbf{Observation Period}} & \multirow{2}{*}{\textbf{Central $\lambda$}} & \multicolumn{2}{c}{\textbf{N$_{\mathrm{obs}}$}} \\
\cline{5-6}
 & & & & \textbf{Spectroscopy} & \textbf{Photometry} \\
 (1)&(2)&(3)&(4)&(5)&(6)\\
\hline
\multirow{3}{*}{PKS~1510-089}
 & Bok 2.3\,m \& Kuiper 1.54\,m & 2008-2017 & 5500 \AA & 271 & 363 \\
 & SMARTS 1.3\,m (V band)   & 2009-2015 & 5510 \AA & --  & 441 \\
 & Metsähovi radio telescope 13.7\,m (37\,GHz)     & 2009-2015 & 200$\mu$m & -- & 341 \\
\hline
\multirow{2}{*}{PKS~0736+017}
 & Bok 2.3\,m \& Kuiper 1.54\,m & 2014-2018 & 5500 \AA & 127 & 133 \\
& ASAS-SN (V-band)& 2015-2018 & 5510 \AA & -- & 58 \\
 & Metsähovi radio telescope 13.7\,m (37\,GHz)               & 2014-2018 & 200$\mu$m & -- & 85 \\
\hline
\end{tabular}
\label{tab:obs_summary}
\begin{flushleft}
\textbf{Note:} Column (1): Source, (2): Telescope/Instrument, (3): Observation period, (4): Central wavelength, (5) and (6): Number of spectroscopy and photometry observations, respectively. Radio observation are from 13.7-m radio telescope monitoring at 37 GHz.
\end{flushleft}
\end{table*}
\subsection{Reverberation mapping Optical data}
The V-band continuum and emission line light curves for \textit{PKS1510} and \textit{PKS0736} used in this study were directly taken from R2020 and P2022, respectively. Below is a concise summary of the observational procedures, data sources, and reduction steps as described in these works. Additionally, the detailed description of the optical and radio data for both of these sources is also mentioned in the Table. \ref{tab:obs_summary}.

The V-band continuum photometry and optical spectra were primarily sourced from the publicly available Steward Observatory (SO) monitoring program\footnote{\url{https://james.as.arizona.edu/~psmith/Fermi/DATA/individual.html}}, which is part of a broader spectropolarimetric effort in support of the \textit{Fermi} Gamma-Ray Space Telescope mission \citep{Smith2009}. Observations were conducted using the 2.3m Bok Telescope (Kitt Peak) and the 1.54m Kuiper Telescope (Mount Bigelow), both equipped with the SPOL spectropolarimeter \citep{1992ApJ...398L..57S}. This instrument employs a 600 mm$^{-1}$ grating, covering 4000-7550\AA with 4\AA/pixel dispersion; typical spectral resolution ranges from 16-24\AA depending on slit width. For full details of the observing setup and data reduction, see \citet{Smith2009}. Briefly, differential photometry was performed using a comparison field star, with instrumental magnitudes for both AGN and comparison star extracted using a synthetic Johnson V-bandpass applied to the spectroscopic data. Flux calibration relied on average sensitivity curves derived from spectrophotometric standard stars, with a final nightly rescaling to match the synthetic V-band photometry \citep{Smith2009}.

For \textit{PKS1510}, R2020 analyzed 363 V-band photometric points (Dec 2008–July 2017) and 341 flux-calibrated optical spectra (Dec 2008–June 2017). They have corrected each spectrum for Galactic extinction \citep{Schlegel1998}, shifted to the rest frame (z=0.361), and modeled with a power-law continuum plus an Fe\textsc{ii} template \citep{Fitzpatrick1999}, masking out emission lines. Residual spectra were fit with Gauss-Hermite and Gaussian profiles to extract the H$\beta$ and [O,\textsc{iii}] emission lines (4740-5050\AA), imposing FWHM constraints. Parameter uncertainties were quantified via Monte Carlo simulations. After removing 70 low S/N spectra, they generated an H$\beta$ lightcurve with 271 spectra. In addition, V-band data from the Small and Moderate Aperture Research Telescope System (SMARTS) monitoring program \citep{Bonning2012} were utilised, based on observations made with the 1.3m telescope at the Cerro Tololo Inter-American Observatory. Lastly, the onset of the flaring state beyond MJD 57100 is clearly evident in the $\gamma$-ray flux, as shown in the first panel of Fig. 2 shown in R2020, and was therefore omitted from our dataset to avoid contamination from high jet activity episodes. Consequently, we restricted our analysis to the quiescent interval between MJD 55000 and 57100. Finally, we binned the data into 3-day intervals to suppress short-timescale photometric noise and mitigate the impact of irregular sampling, while preserving the intrinsic variability on timescales relevant for damped random walk (DRW) modeling.

For \textit{PKS0736}, P2022 analyzed 133 V-band photometric points (Nov 2014–May 2018), and 127 flux-calibrated spectra from the SO database and constructed an H$\beta$ emission light curve using a direct integration method. After rest-frame correction, a local power-law of the form $P(\lambda) = \alpha \lambda^{\beta}$ was fitted to continuum regions adjacent to the H$\beta$+[O\textsc{iii}] complex. The continuum-subtracted spectra were then integrated within fixed wavelength intervals to measure emission line fluxes. The $5100\,\text{\AA}$ continuum flux was measured from the best-fit power-law. To account for slit loss, seeing, and calibration uncertainties, the spectra were rescaled using the [O\textsc{iii}] $\lambda$5007 line, which was assumed to be non-variable over the monitoring period. Reported flux uncertainties include both Poisson noise and systematics, with the latter estimated via the median-filter approach described by \citet{Zhang2019}. Similar to PKS1510, evidence of a flaring state before MJD 57285 is apparent in the $\gamma$-ray flux for PKS0736, as shown in Fig.~2 of P2022, and this period was excluded to prevent contamination from high-activity phases. The analysis was therefore carried out using only the quiescent interval between MJD 57285 and 58250. Additionally, we also combined V-band photometric data points (centered at 4747{\AA}) from the All-Sky Automated Survey for Supernovae \citep[ASAS-SN;][]{Shappee2014, Kochanek2017, Hart2023} to improve the cadence and increase the temporal coverage. Finally, we intercalibrated the SO and ASAS-SN photometric data points using the {\tt PyCALI} software \citep{Pycali2014} with SO photometry V-band as the reference band.

\subsection{Polarization data}
We use optical polarization data for PKS0736 and PKS1510 obtained from the SO spectropolarimetric monitoring program. The dataset provides measurements of the Stokes parameters $Q$ and $U$, from which the fractional linear polarization ($p$) and polarization position angle ($\theta$) are derived. The polarization values are corrected for statistical bias following the prescription of \citet{wardle1974}, and uncertainties are estimated based on photon statistics. No correction for interstellar polarization has been applied. The polarization position angle $\theta$ (in degrees, measured east of north) is calculated using $\theta = 0.5 \tan^{-1}(U/Q)$, with its uncertainty estimated as $\sigma_\theta = 28.65$, $\sigma_p/p$ for $p/\sigma_p > 3$. 

\subsection{Radio data}
The 37~GHz radio observations for the blazars PKS~0736+017 and PKS~1510$-$089 were obtained using the 13.7~m diameter Aalto University Mets\"{a}hovi radio telescope in Finland. The telescope is equipped with a HEMPT (high electron mobility pseudomorphic transistor) front end operating at room temperature. The measurements were taken using Dicke switched ON--ON observations, alternating the source and the sky in each feed horn, with typical integration times ranging between 1200 and 1800~s per flux density data point. 

The detection limit of the telescope at 37~GHz is approximately 0.2~Jy under optimal conditions, and data points with a signal-to-noise ratio $< 4$ are handled as non-detections. A comprehensive description of the data reduction and analysis procedures can be found in \citet{Terasranta1998}. The final error estimates in the flux densities incorporate both the measurement root-mean-square (RMS) and the uncertainty associated with the absolute calibration.
\section{Evidence for Jet Contamination}
\label{sec:Jet evidence}
Broadband emission in blazars, particularly in FSRQs, is often significantly contaminated by Doppler-boosted non-thermal emission from relativistic jets. This jet contribution can bias measurements of the thermal continuum, distort reverberation signals, and bias luminosity-dependent estimates of BLR size and black hole mass. Here, we evaluate the evidence for jet contamination in  PKS0736 and PKS1510 based on spectral behavior, light curve comparisons, and non-thermal dominance parameters.

\subsection{Non-Thermal Dominance (NTD) Parameter}
The non-thermal dominance (NTD) parameter, defined as the ratio of the observed continuum luminosity to the luminosity predicted from the emission-line strength (typically H$\beta$), provides a diagnostic of jet contamination \citep{Shaw2012}. In the absence of strong jet influence, the continuum and line luminosities are expected to maintain a nearly linear correlation.

For PKS1510, R2020 computed the NTD values by comparing the measured optical continuum luminosity at $5100\,\text{\AA}$ to the value inferred from the H$\beta$ line luminosity using the empirical relation from R2020. The NTD exceeded 2.0 during high-activity periods, particularly in epochs coinciding with strong $\gamma$-ray flares observed by \textit{Fermi}. They also noted that the broad-line luminosity varied significantly less than the continuum, reinforcing the conclusion that jet variability dominates the optical continuum during flaring episodes. For PKS0736, P2022 found that during $\gamma$-ray flaring states, the NTD increased to values as high as $\sim$2.3, suggesting that more than half of the optical continuum could be attributed to non-thermal jet emission. Even during quiescent states, the NTD remained above unity, indicating a persistent jet contribution to the optical continuum.
\subsection{Line versus Continuum Variability}
\begin{figure*}
    \includegraphics[height=12cm, width=9cm]{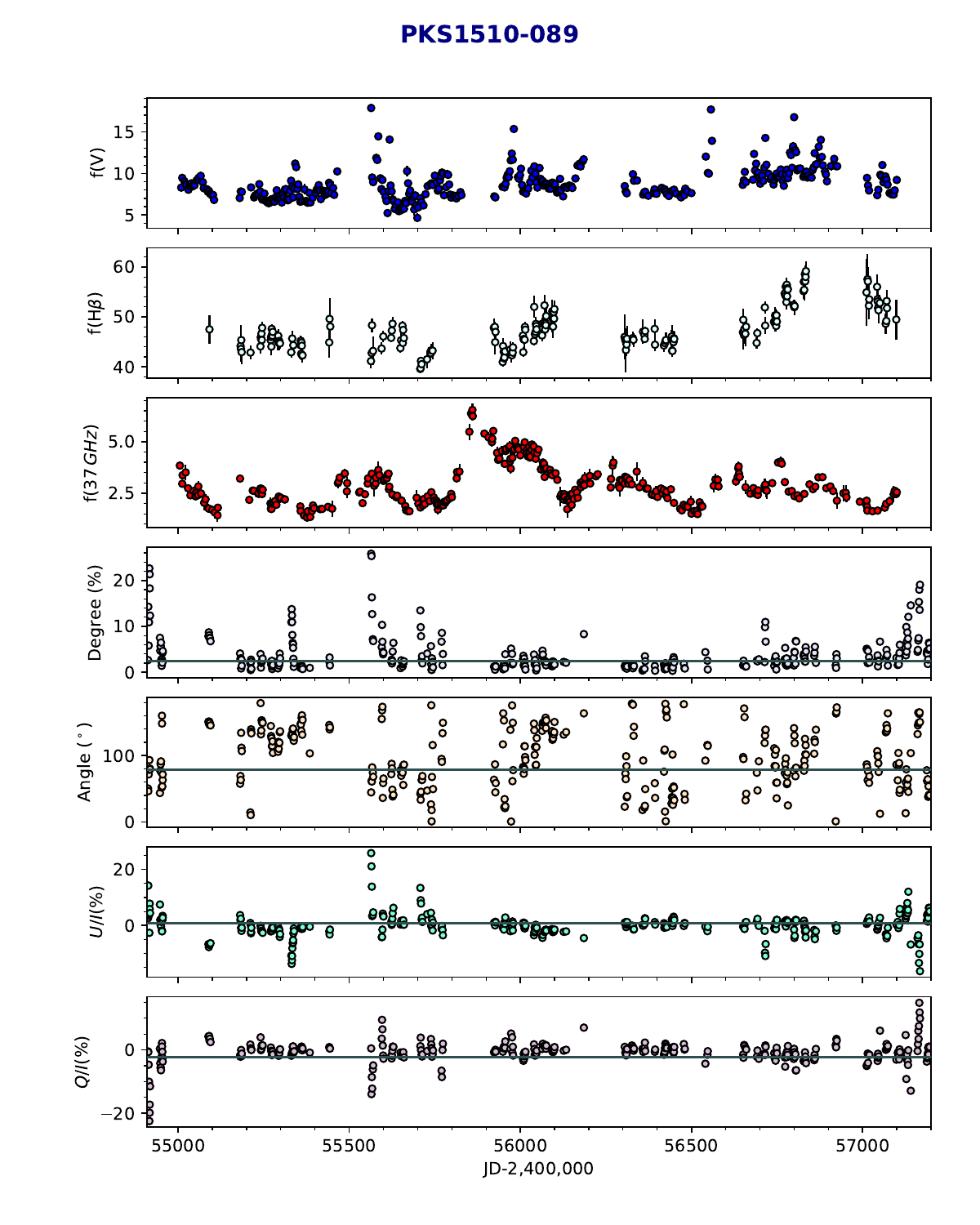}
    \includegraphics[height=12cm, width=9cm]{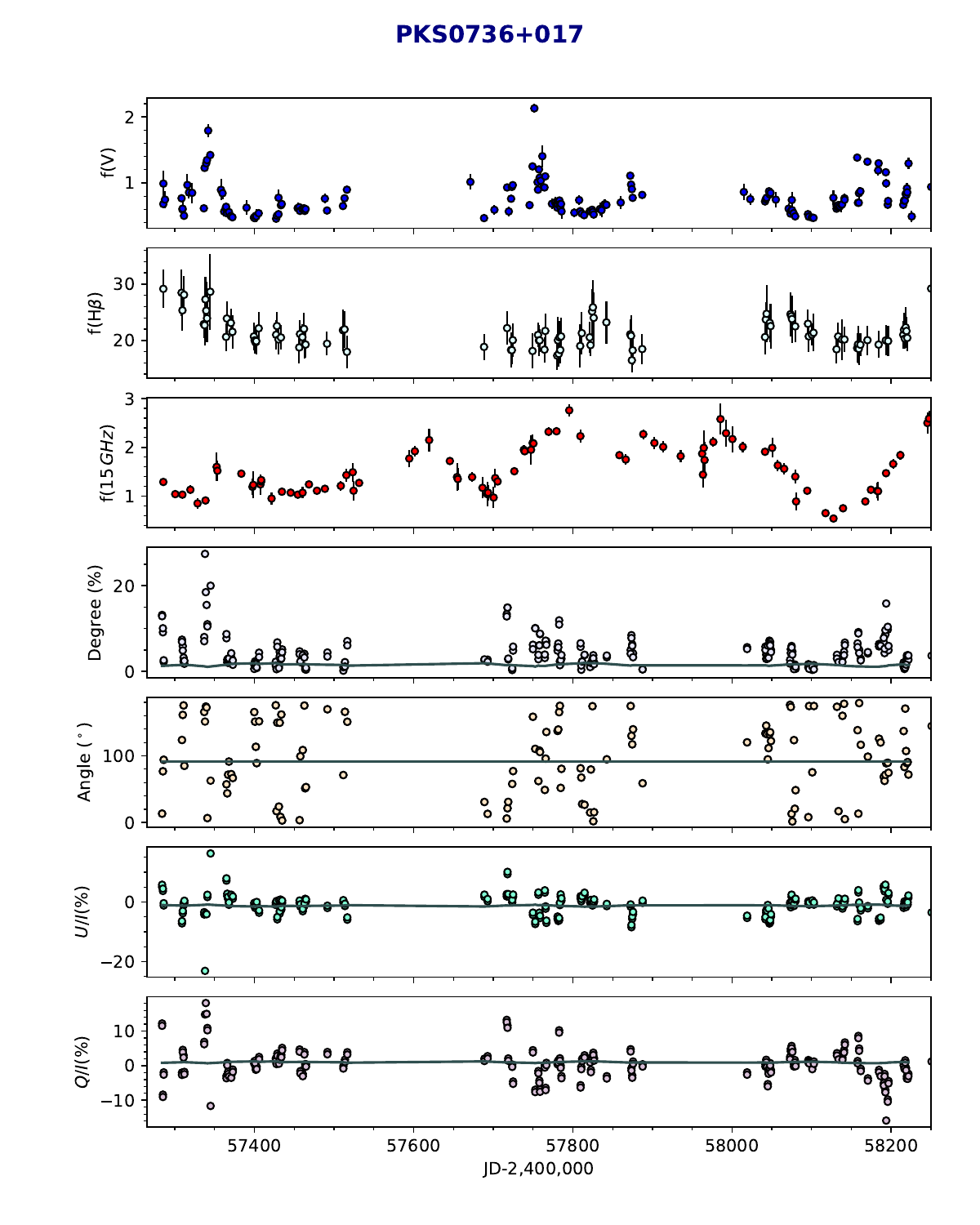}
    \caption{L-R: For PKS~1510-089 and PKS~0736+017, the first panel shows the V-band optical continuum. The second panel show the H$\beta$ emission line directly taken from R2020 and P2022. The third shows the 37 GHz radio light curve from 13.7~m diameter Aalto University Mets\"{a}hovi radio telescope. The fourth panel represents degree of polarization (\%) with observed data and model fitted with DECOMP, representing the long-term polarization trend. The last three panels show the polarization angle ($^{\circ}$), Stokes parameter ratio U/I(\%), and Q/I (\%), respectively.}
    \label{fig:Fitted_polar}
\end{figure*}
A key diagnostic for jet contamination is the decoupling between line and continuum variations. In a thermally dominated AGN, the broad-line light curve is expected to closely follow the continuum with a time delay corresponding to the BLR light-travel time. However, in both PKS0736 and PKS1510, the continuum exhibited stronger and more rapid variability compared to the broad H$\beta$ line. P2022 showed that while the continuum light curve of PKS0736 displayed sharp variability features, the H$\beta$ line response was smoother and showed less variability power. This mismatch suggests that a significant portion of the variable continuum arises from the jet and does not contribute to photoionization. R2020 made similar observations in PKS1510, noting that some continuum outbursts had no corresponding line response.

\subsection{Deviations from the R–L Relation}
Another indication of jet contamination of these radio-loud objects is a systematic deviation from the empirical $R_{\mathrm{BLR}}$–$L_{5100}$ relation, which is calibrated using thermally dominated AGNs \citep{Bentz2013, Woo2024, Shen2024}. In both PKS0736 and PKS1510, the measured BLR lags are lower than the values predicted by their observed continuum luminosities.

In the case of PKS0736, the H$\beta$ lag of $\sim$66 days corresponds to a slightly smaller BLR size than 87 light-days expected from its $5100\,\text{\AA}$ luminosity \citep{Bentz2013}. After correcting the luminosity using the NTD parameter, the source aligns more closely with the standard relation. A similar offset is seen for PKS1510, whose measured H$\beta$ lag of $\sim$45 days falls short of three times the value predicted using its observed continuum. These deviations further support the scenario that jet contamination artificially inflates the observed luminosity, leading to overestimated BLR sizes if uncorrected.
\subsection{Variation in polarization data}
As also noted in Li2020, the most reliable evidence for jet contamination in the optical continuum via synchrotron emission in both sources is the variability observed in the polarization data, as shown in Fig.~\ref{fig:Fitted_polar}. For PKS0736, the polarization degree in the fourth panel of Fig.~\ref{fig:Fitted_polar} exhibits significant variability, although the median remains close to 2-3\%. The fifth panel shows the polarization angle, while the sixth and seventh panels display the variations in the ratios of the Stokes parameters U/I (\%) and Q/I (\%), respectively. In contrast, PKS1510 exhibits a higher median polarization degree and a less variable polarization angle. The relatively low median polarization degree suggests a substantial contribution from weakly polarized thermal disk emission, while the observed variability indicates an additional synchrotron component associated with the jet.
\section{Separating Disk and Jet Contributions}
\label{sec:DECOMP}
The optical emission C$_{t}$(t) that is observed is a collective contribution of continuum emission coming from the Disk C$_{d}$(t) and the Jet C$_{j}$(t), such as
\begin{align}
\mathrm{C_{t}(t)}=\mathrm{C_{d}(t)} +{\mathrm{C_{j}(t)}}, 
\end{align}
where the subscripts "t", "d", and "j" refer to the total, disk, and jet emissions, respectively. To disentangle the thermal and nonthermal components contributing to the total observed optical emission, we employ the \texttt{DECOMP} algorithm developed by Li2020. In this section, we briefly describe the methodology and implementation of \texttt{DECOMP}, emphasizing its effectiveness in separating the disk and jet contributions in blazar spectra. \texttt{DECOMP} performs a Bayesian decomposition of the observed optical variability by modeling the disk and jet components as independent Damped Random Walk (DRW) processes. Eq.~\ref{Eq:DRW} represents the auto-covariance function for C$_{\mathrm{d}}$ and C$_{\mathrm{j}}$, characterized by their respective damping timescales ($T_d$ and $T_j$) and long-term standard deviations ($\sigma_d$ and $\sigma_j$). 
\begin{align}
S_{\mathrm{\star}}(\Delta t) = \langle C_\star(t_k) C_\star(t_m) \rangle = \sigma_\star^2 \exp\left( -\frac{|\Delta t|}{T_\star} \right),
\label{Eq:DRW}
\end{align}
Here, $\star$ is for disk (d) or jet (j). The most important assumption is that over a short timescale, the variability of C$_{d}$(t) and C$_{j}$(t) is independent and not correlated.
\begin{table*}[t]
\centering
\caption{Free Parameters and Source-Specific Ranges}
\label{tab:free_parameters}
\begin{tabular}{lcccl}
\hline
\textbf{Parameter} & \textbf{Prior} & \textbf{PKS~1510-089} & \textbf{PKS~0736+017} & \textbf{Definition} \\
\hline
$\sigma_d$        & Logarithmic & ($10^{-3}$, 0.8)       & ($10^{-3}$, 1)         & Long-term standard deviation of $C_{\rm disk}$ \\
$T_d$             & Logarithmic & (1, $10^{5}$)          & (1, $10^{5}$)          & Characteristic damping timescale of $C_{\rm disk}$ (days) \\
$\sigma_j$        & Logarithmic & ($10^{-5}$, $10^{-1}$) & ($10^{-3}$, 0.8)       & Long-term standard deviation of $C_{\rm jet}$ \\
$T_j$             & Logarithmic & (1, $10^{5}$)          & (1, $10^{5}$)          & Characteristic damping timescale of $C_{\rm jet}$ (days) \\
$f_{H\beta}$      & Logarithmic & ($10^{-3}$, $10^{3}$)  & ($10^{-3}$, $10^{3}$)  & Amplitude of the TF for the H$\beta$ light curve \\
$\tau_{H\beta}$   & Uniform     & (1, 140)               & (0, 300)               & Mean time delay of the TF for the H$\beta$ light curve (days) \\
$\omega_{H\beta}$ & Logarithmic & (0, $10^{3}$)          & (0, $10^{3}$)          & Standard deviation of the TF for the H$\beta$ light curve (days) \\
$f_j$             & Logarithmic & ($10^{-3}$, $10^{3}$)  & ($10^{-3}$, $10^{3}$)  & Amplitude of the TF for the 37~GHz light curve \\
$\tau_j$          & Uniform     & (0, 150)               & (0, 500)               & Mean time delay of the TF for the 37~GHz light curve (days) \\
$\omega_j$        & Logarithmic & (1, $10^{3}$)          & (1, $10^{3}$)          & Standard deviation of the TF for the 37~GHz light curve (days) \\
$f_d$             & Uniform     & (0, 1)                 & (0, 1)                 & Ratio of the mean of $C_{\rm disk}$ to that of $C_{\rm total}$ \\
$\bar{p}_j$       & Uniform     & (0, 6)\%               & (0, 2.5)\%             & Average polarization degree of $C_{\rm jet}$ \\
$\theta_j$        & Uniform     & (0, 180$^\circ$)       & (0, 180$^\circ$)       & Average polarization angle of $C_{\rm jet}$ \\
\hline
\end{tabular}
\begin{flushleft}
\textbf{Note.} Prior types and ranges of the free parameters for PKS~1510-089 and PKS~0736+017. The transfer function (TF) is assumed to be Gaussian.
\end{flushleft}
\end{table*}
Assuming the perturbations propagate along the jet from denser to less dense regions, the emitted photon energy gradually decreases from $\gamma$/X-ray to UV/optical to radio wavelengths \citep{Marscher1985}. Similarly, the H$\beta$ variations are reflections of the thermal fluctuations originating from the accretion disk. Consequently, the jet continuum correlates with the radio emission, while the disk continuum correlates with the H$\beta$ emission line light curve. These connections are described through a transfer function, as shown in Eq.~\ref{eq:convolution}. The radio emission represents a lagged version of the jet continuum contribution, resulting from its convolution with the transfer function, which encodes the kinematic and dynamical information of the jet. Likewise, the H$\beta$ emission line light curve is a lagged counterpart of the disk continuum contribution within the total optical emission.
\begin{align}
F_{*}(t) = \int \Psi_{\mathrm{\star}}(\tau) C_{\mathrm{\star}}(t - \tau) \, d\tau,
\label{eq:convolution}
\end{align}
where $\star$ represents either the H$\beta$ emission line or the radio emission, $\psi$ is the transfer function that characterizes the geometry and response structure of the emission region, and C$_{\mathrm{cont}}$ is the ionizing continuum. The adopted transfer function is Gaussian in nature, as shown in Eq.~\ref{Eq:TF}, and is defined by parameters such as amplitude ($f$), time delay ($\tau$), and standard deviation ($\omega$).
\begin{equation}
\Psi(\tau) = \frac{f_{\mathrm{\star}}}{\sqrt{2\pi \omega_{\mathrm{\star}}^2}} \exp\left(-\frac{(\tau - \tau_{\mathrm{\star}})^2}{2 \omega_{\mathrm{\star}}^2}\right),
\label{Eq:TF}
\end{equation}
The observed data fed into the code include the optical V-band continuum light curve (C$_{\mathrm{t}}$(t)), the H$\beta$ emission line light curve (L$_{\mathrm{H_{\beta}}}$(t)), and the 37 GHz radio emission light curve (R$_{\mathrm{j}}$(t)). The parameter priors used for constructing the posterior distribution through Bayesian inference are listed in Table~\ref{tab:free_parameters}. Except for $\tau_{H_{\beta}}$ and $\tau_{j}$, which follow uniform distributions, all other priors are assumed to be logarithmic. As described in Section~4.2 of Li2020, the posterior probability distribution is explored using a diffusive nested sampling technique\footnote{CDNest: \url{https://github.com/LiyrAstroph/CDNest}}, based on the method introduced by \citet{Brewer2011}. This technique enables efficient sampling of complex posterior landscapes and yields robust estimates along with associated uncertainties for the model parameters. By incorporating polarization data specifically, the polarization degree ($p$), polarization or position angle ($\theta$), and Stokes parameters $U$, $Q$, $I$, and $V$ (where $V = 0$ for linear polarization) along with an additional free parameter $f_{\mathrm{q}}$, defined as the ratio of mean disk emission C$_{\mathrm{d}}$(t) to the total optical continuum emission C$_{\mathrm{t}}$(t), the model decomposes the mean disk and jet components. The observed optical polarization for both sources is expressed as follows:
\begin{align}
I = C_t = C_d + C_j, 
\label{eq:StokesI}
\end{align}
\begin{equation}
Q = C_d p_d \cos(2\theta_d) + C_j p_j \cos(2\theta_j) + C_t p_* \cos(2\theta_*), 
\label{eq:StokesQ}
\end{equation}
\begin{equation}
U = C_d p_d \sin(2\theta_d) + C_j p_j \sin(2\theta_j) + C_t p_* \sin(2\theta_*), 
\label{eq:StokesU}
\end{equation}
where subscripts "d", "j", and "$\star$" represent disk, jet, and interstellar polarization, respectively.
As described in Li2020, the observed polarization is a combination of interstellar polarization and polarization from disk and jet emissions. Unlike interstellar polarization, which remains constant, the disk and jet polarization vary with time.
\begin{equation}
p = \frac{\sqrt{U^2 + Q^2}}{I}, 
\label{eq:poldegree}
\end{equation}
\begin{equation}
\theta = \frac{1}{2} \tan^{-1} \left( \frac{U}{Q} \right).
\label{eq:poltheta}
\end{equation}
The parameter $f_{\mathrm{q}}$ is primarily constrained by fitting the polarization data, namely $U$, $Q$, $p$, and $\theta$ as described in Eq.~\ref{eq:StokesI}, \ref{eq:StokesQ}, \ref{eq:StokesU}, \ref{eq:poldegree}, and \ref{eq:poltheta}. The model fitting with the observed data is illustrated in Fig.~\ref{fig:Fitted_polar}. The interstellar polarization degree can be estimated using the approximation $p_{\mathrm{\star}}(\%) = 9E(B{-}V)$ \citep{Serkowski1975}. For PKS0736 ($z \approx 0.189$), the line-of-sight reddening is $E(B{-}V) \approx 0.121$~mag, based on the \citet{Wu1977} recalibration of the \citet{Schlegel1998} and \citet{Schlafly2011} dust maps. For PKS1510 ($z \approx 0.361$), the recalibrated reddening is $E(B{-}V) \approx 0.085$~mag, slightly lower than the original SFD estimate of $\approx 0.097$~mag. These values suggest moderate to low levels of interstellar polarization contamination for both sources. For the interstellar polarization angle, we adopt $\theta_{\mathrm{\star}} = 32^{\circ}$ with $p_{\mathrm{\star}}(\%) = 0.77$ for PKS1510 and $\theta_{\mathrm{\star}} = 1^{\circ}$ with p$_{\mathrm{\star}}(\%) = 1.10$ for PKS0736. Since there were no direct interstellar polarization angle measurements for these sources, we have chosen the closest source star from Heiles Stellar Polarization Catalogs \citep{Heiles2000}. We assume the disk polarization is negligible compared to the jet, thus setting $p_{\mathrm{d}} = 0$, and consider $\theta_{\mathrm{d}}$ to be insignificant. The priors for $f_{\mathrm{q}}$, $\bar{p_{\mathrm{j}}}$, and $\theta_{\mathrm{j}}$ are listed in Table~\ref{tab:free_parameters}. Upon completion of the run, the total continuum is decomposed into disk and jet components, and the posterior distributions of the model parameters are presented in Table~\ref{tab:inferred_params}.
\section{Results}
\label{sec:results}
\begin{table}
\caption{Posterior Parameter Values based on \texttt{DECOMP} modeling}
\label{tab:inferred_params}
\begin{tabular}{lcc}
\hline
\textbf{Parameter} & \textbf{PKS~1510-089} & \textbf{PKS~0736+017} \\
\hline
$\log \sigma_d$      & $-1.31^{+0.05}_{-0.06}$ & $-0.97^{+0.12}_{-0.53}$\\
$\log (T_d/\mathrm{day})$   &$1.26^{+0.12}_{-0.11}$ & $0.97^{+0.83}_{-0.24}$\\
$\log \sigma_j$      & $-2.41^{+0.10}_{-0.11}$ & $-1.14^{+0.30}_{-0.96}$\\
$\log (T_j/\mathrm{day})$   &$2.31^{+0.47}_{-0.24}$  & $0.74^{+0.34}_{-0.28}$\\
$\log f_{H\beta}$     & $-0.03^{+0.01}_{-0.01}$ & $-0.12^{+0.08}_{-0.09}$ \\
$\tau_{H\beta}$ /day &$54.4^{+4.4}_{-5.4}$  & $84.8^{+23.2}_{-18.4}$\\
$\log (\omega_{H\beta}/\mathrm{day})$ & $1.64^{+0.04}_{-0.03}$ &$1.57^{+0.36}_{-0.40}$ \\
$\log f_j$           & $0.85^{+0.12}_{-0.08}$ & $0.77^{+0.52}_{-0.20}$\\
$\tau_j$/day         & $8.3^{+2.5}_{-1.4}$ & $41.6^{+18.5}_{-11.5}$ \\
$\log (\omega_j/\mathrm{day})$ & $0.17^{+0.04}_{-0.12}$ &$1.50^{+0.07}_{-0.11}$\\
$f_d$                & $0.37^{+0.03}_{-0.03}$ & $0.70^{+0.01}_{-0.01}$\\
$\bar{p_j}$\%             & $3.60^{+0.03}_{-0.03}$ &$2.05^{+0.03}_{-0.02}$\\
$\theta_j$         & $78.8^{+0.1}_{-0.1}$ &$91.8^{+0.5}_{-0.5}$\\
\hline
\end{tabular}
\\[1ex]
\begin{flushleft}
    \textbf{Note.} Parameter values are determined from the medians of the posterior probability distributions, and the uncertainties represent the 68.3\% confidence intervals. The values of $f_d$, $\bar{p_j}$, and $\theta_j$ are given by fixing the interstellar polarization degree and angle for PKS1510 at 0.77\% with 32 degrees and 1.10\% with 1 degree for PKS0736.
\end{flushleft}
\end{table}
\subsection{Overview}
\begin{figure*}
\centering
    \includegraphics[scale=0.4]{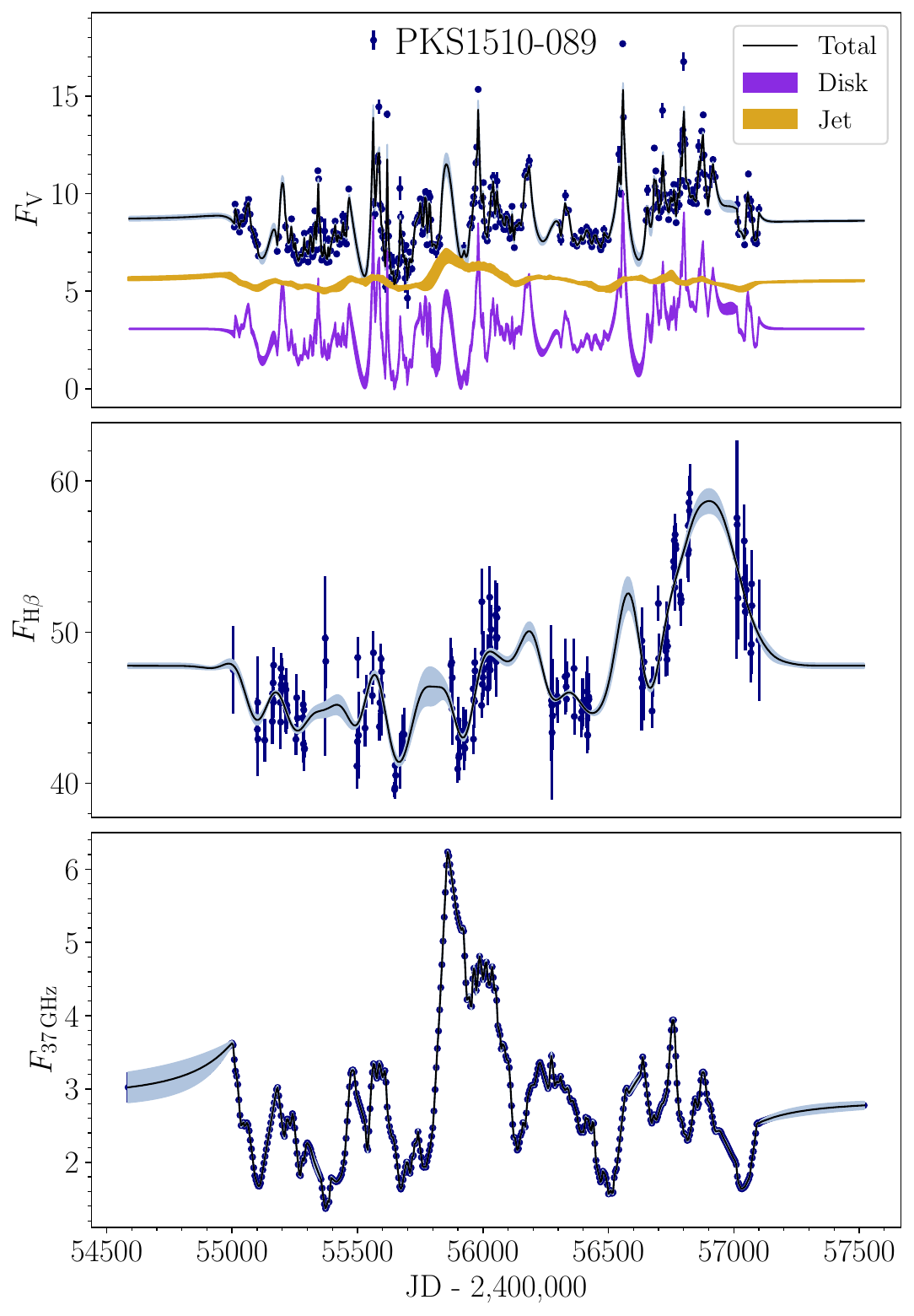}
    \includegraphics[scale=0.4]{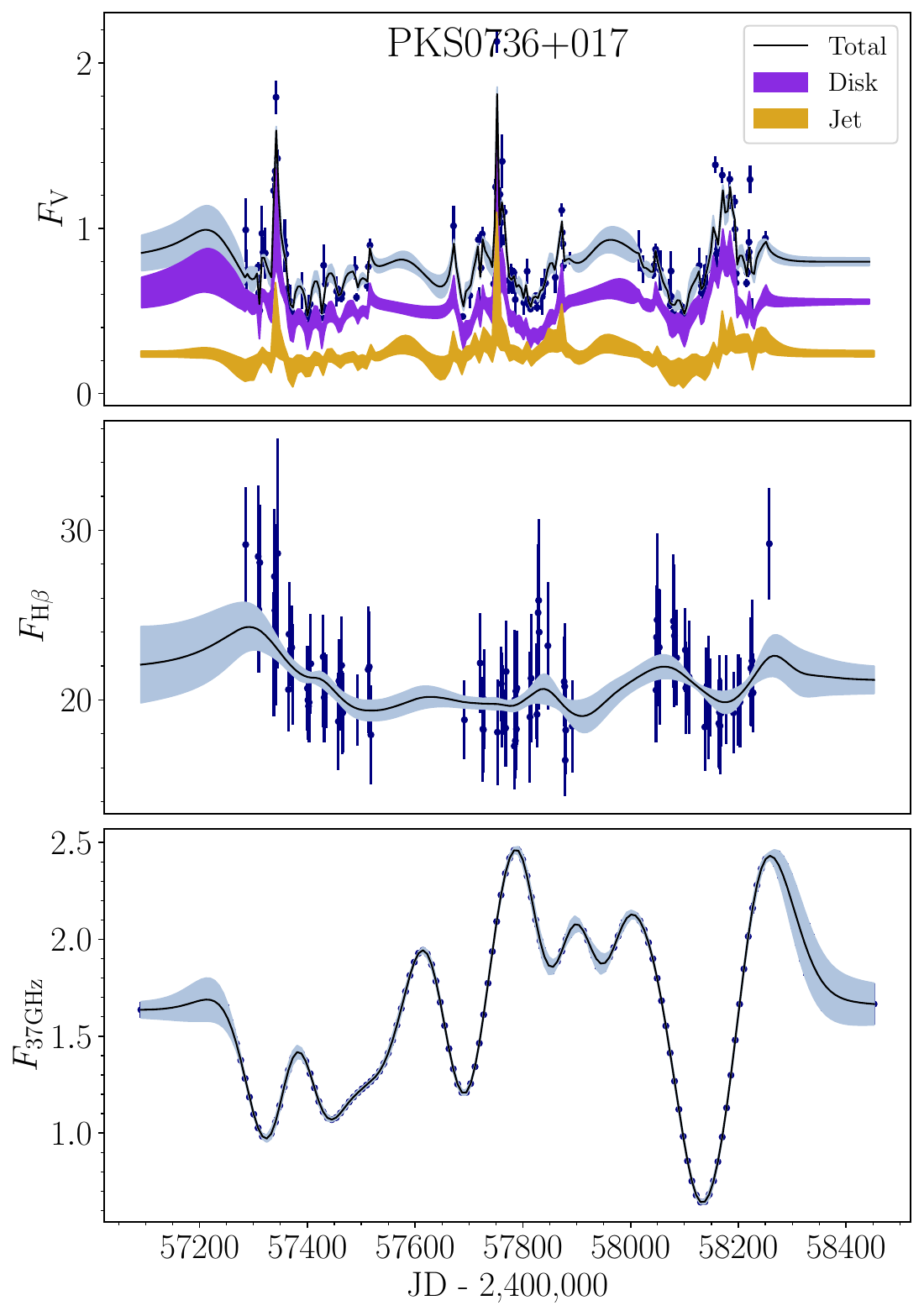}
    \caption{L-R: Light curve reconstruction from \texttt{DECOMP} modelling for PKS~1510-089 and PKS~0736+017. The first panel shows the reconstructed total optical continuum over observed data, decomposed disk and jet emissions. The second and third panels show the reconstructed H$\beta$ emission line flux and 37 GHz Radio data, respectively, overlaid on the observed data. }
    \label{fig:LC_recon}
\end{figure*}
In the present study, we adopt a similar approach to that used for 3C 273 by Li2020 to analyse the polarization light curves of PKS0736 and PKS1510, obtained from the SO monitoring program. Both sources exhibit variability in optical polarization throughout the monitoring period, making them suitable for applying the \texttt{DECOMP} method. By isolating the thermal component of the optical continuum, we aim to more accurately constrain the BLR response and evaluate the impact of jet contamination on RM measurements in these FSRQs.

\begin{figure*}
    \centering
    \includegraphics[scale=1,width=1\linewidth]{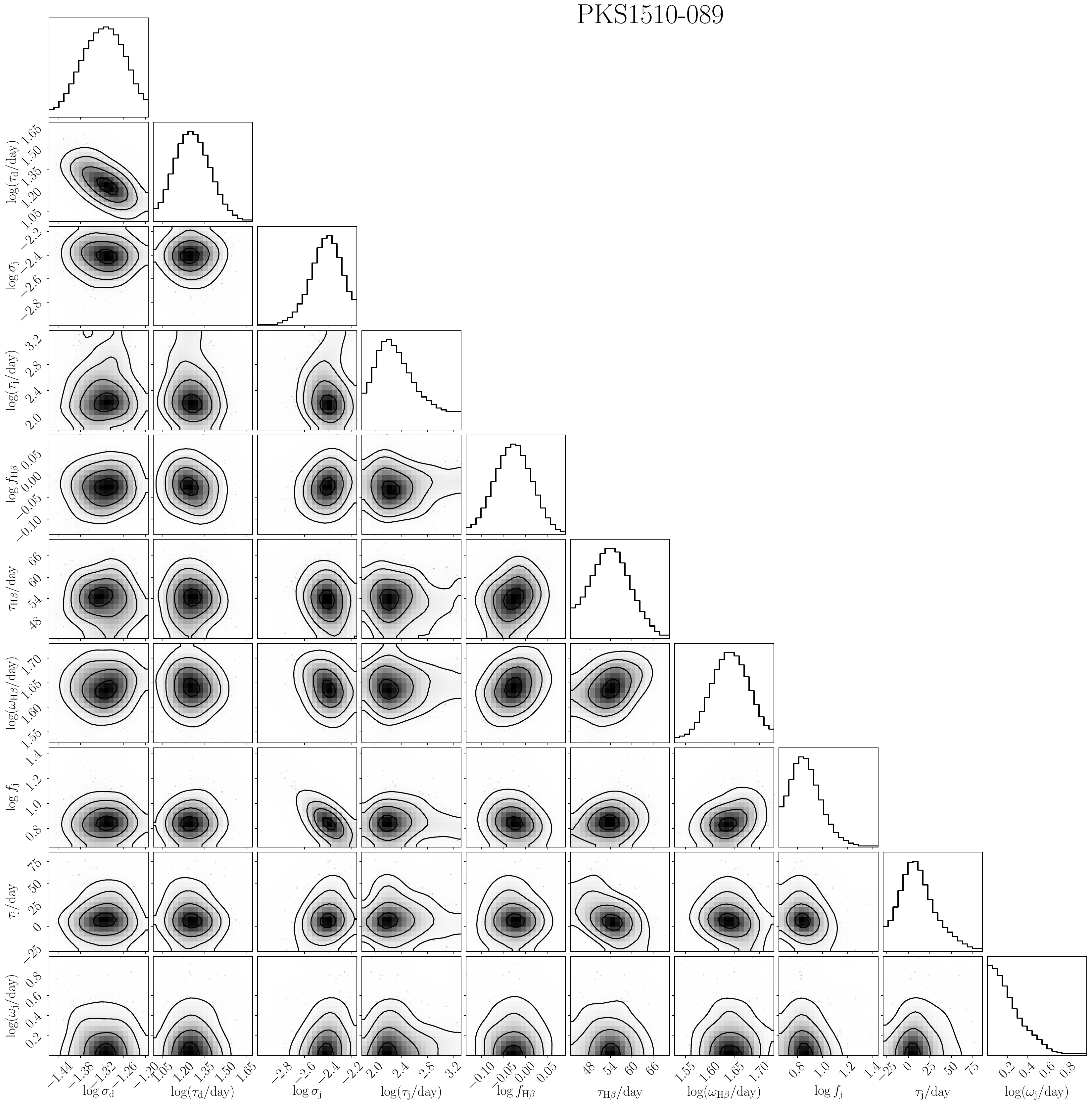}
    \caption{Posterior distributions of the free parameters from the \texttt{DECOMP} modeling for PKS~1510-089 in 1- and 2-dimension.The distributions illustrate the likelihoods and covariances between parameters such as disk flux, jet flux, and fractional contributions, obtained using diffusive nested sampling. The contours are plotted at 1$\sigma$, 1.5$\sigma$, and 2$\sigma$.}
    \label{fig:Histogram1}
\end{figure*}

\begin{figure*}
    \includegraphics[scale=1,width=1\linewidth]{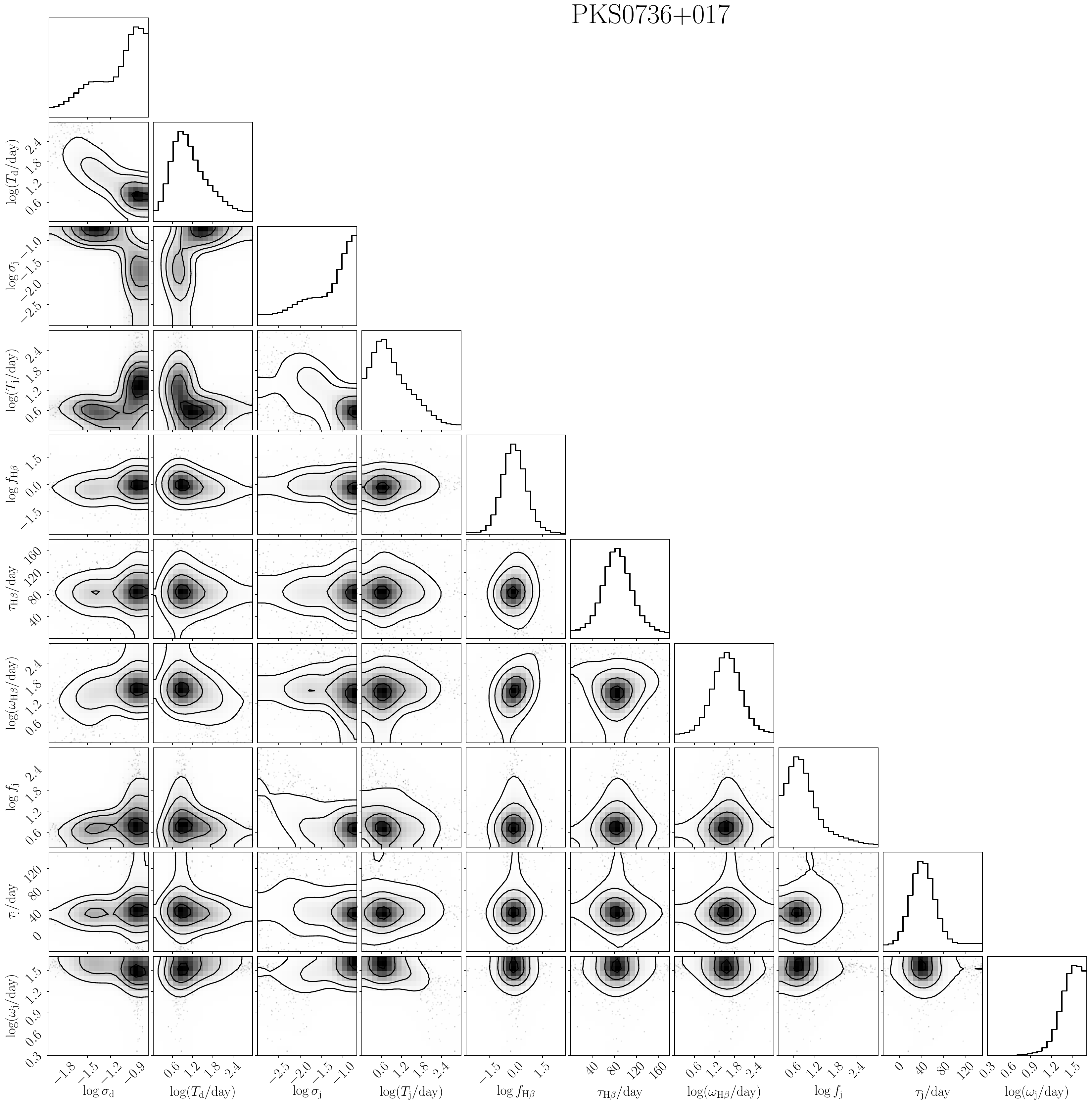}
    \caption{Posterior distributions of the free parameters from the \texttt{DECOMP} modeling for PKS~0736+017 1- and 2-dimension. The contours are plotted at 1$\sigma$, 1.5$\sigma$, and 2$\sigma$.}
    \label{fig:Histogram2}
\end{figure*}

\begin{figure*}
    \includegraphics[width=0.5\linewidth]{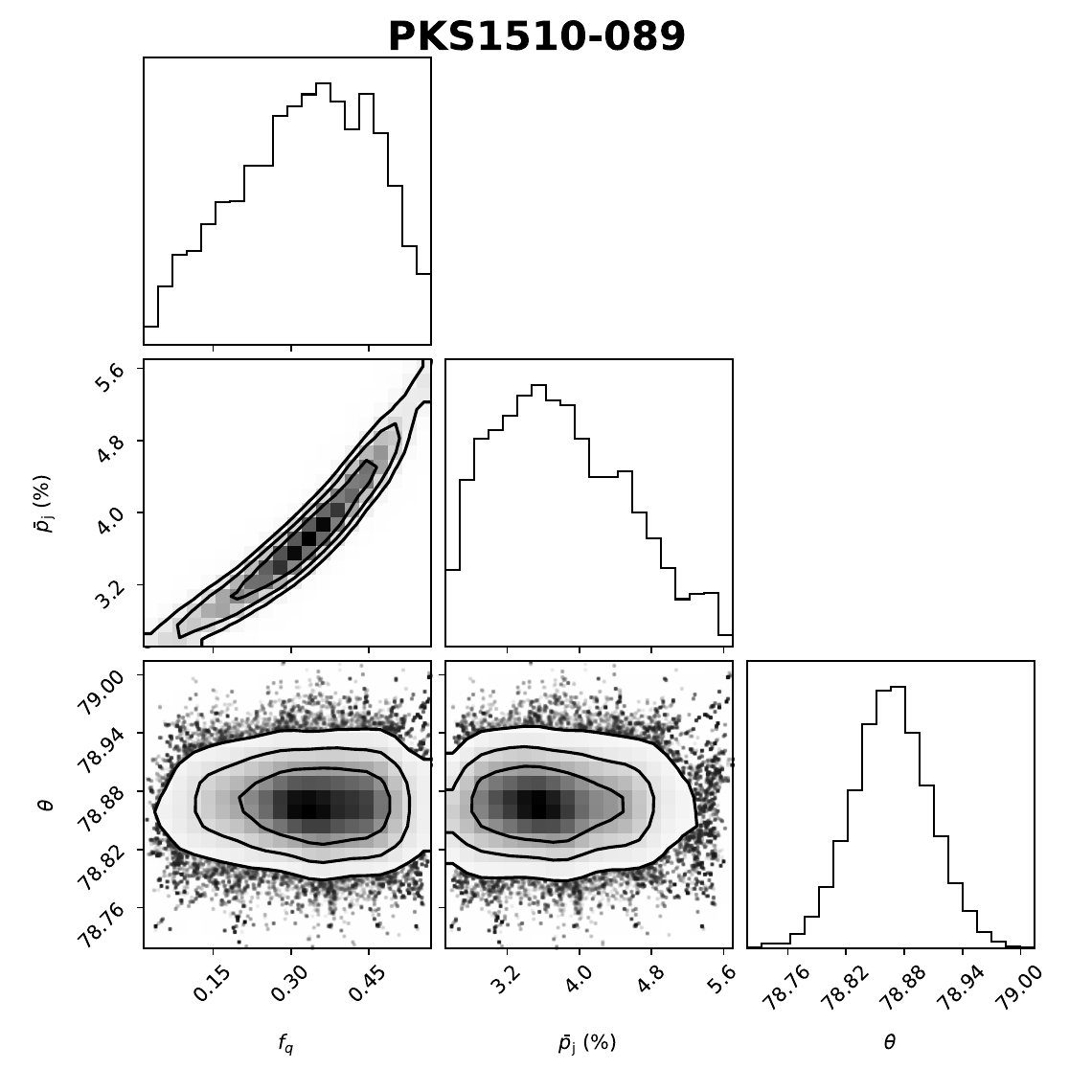}
    \includegraphics[width=0.5\linewidth]{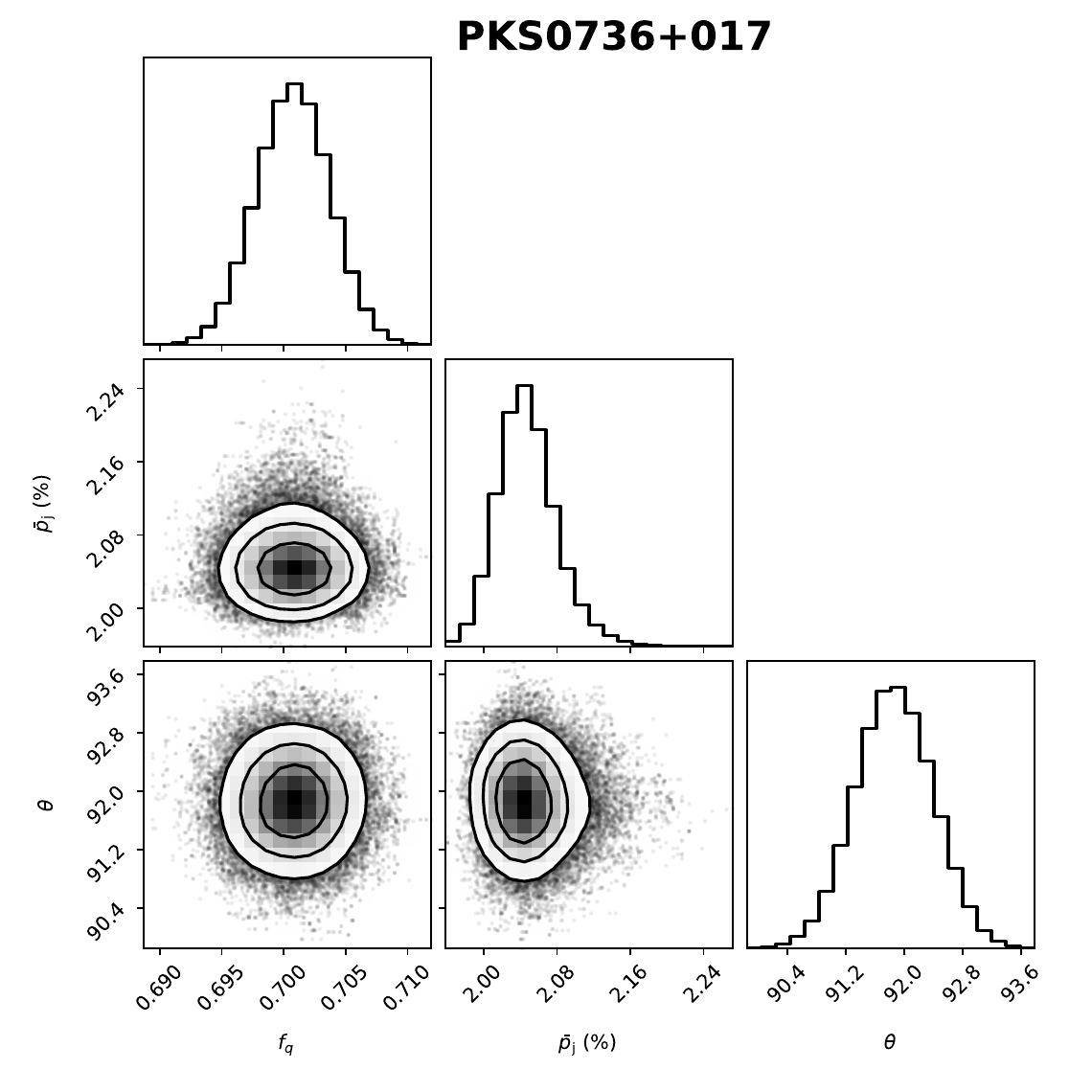}
    \caption{L-R: The Posterior distribution for PKS~1510-089 and PKS~0736+017 for polarization data. These parameters are the ratio of the mean of disk emission to total optical flux f$_{\mathrm{q}}$, average polarization degree ($\bar{p_j}$(\%)), and polarization angle ($\theta$).} 
    \label{fig:Polar_Dist}
\end{figure*}

The physical parameters for the optical disk and jet emissions, as well as those for the transfer functions (TFs) used to model the H$\beta$ and 37 GHz radio light curves, were constrained using the prior ranges listed in Table~\ref{tab:free_parameters}. The polarization-dependent parameters, including the ratio of disk to total optical emission ($f_{\mathrm{d}}$), average polarization degree ($\bar{p_j}$), and polarization angle ($\theta_{\mathrm{j}}$), are obtained by fitting the observed data to the model, as shown in Fig~\ref{fig:Fitted_polar}. The posterior values inferred from the Bayesian analysis are provided in Table~\ref{tab:inferred_params}. The contour plots of the posterior distributions are displayed in Fig~\ref{fig:Histogram1} and \ref{fig:Histogram2}, and the posteriors of the polarization parameters are shown in Fig~\ref{fig:Polar_Dist}. Below, we summarise the key findings for both sources.

For PKS1510, the optical V-band continuum light curve was modeled using a DRW process, with the characteristic damping timescale $\log (T_d/\mathrm{day})$ and long-term standard deviation $\log \sigma_d$ as described in Section~\ref{sec:DECOMP}. Gaussian transfer functions were employed to model the radio and H$\beta$ emission line light curves. The disk variability amplitude $\log \sigma_d$ was tightly constrained to $-1.31^{+0.05}_{-0.06}$, within the prior range (10$^{-3}$, 0.8), and the damping timescale $\log (T_d/\mathrm{day})$ was found to be $1.26^{+0.12}_{-0.11}$. In contrast, the jet emission showed a lower amplitude of $\log \sigma_j = -2.41^{+0.10}_{-0.11}$ and a longer variability timescale of $\log (T_j/\mathrm{day}) = 2.31^{+0.47}_{-0.24}$, consistent with the adopted priors. For the H$\beta$ emission line, the transfer function amplitude $\log f_{H\beta}$ was $-0.03^{+0.01}_{-0.01}$, with a delay $\tau_{H\beta} = 54.4^{+4.4}_{-5.4}$ days. The width of the Gaussian transfer function was $\log (\omega_{H\beta}/\mathrm{day}) = 1.64^{+0.04}_{-0.03}$. For the radio emission light curve, the posterior amplitude was $\log f_j = 0.85^{+0.12}_{-0.08}$, with a delay of $\tau_j = 8.3^{+2.5}_{-1.4}$ days and a width of $\log (\omega_j/\mathrm{day}) = 0.17^{+0.04}_{-0.12}$. The radio emission lag is smaller than H$\beta$ lag, consistent with the finding of lag range 10-50 days by \citet{Yuan2023}, suggesting that the 37GHz emission traces a more compact jet region than the lower frequency radio emission.

For PKS0736, the posteriors for the DRW parameters were similarly obtained. The disk variability amplitude $\log \sigma_d$ was constrained to -0.97$^{+0.12}_{-0.53}$, which lies well within the prior range of (10$^{-3}$, 1). The damping timescale $\log (\tau_d/\mathrm{day})$ was inferred to be 0.97$^{+0.83}_{-0.24}$, consistent with the broad prior range (1.0, 10$^5$). The 37 GHz radio light curve related to jet emission exhibited a variability amplitude of $\log \sigma_j = -1.14^{+0.30}_{-0.96}$ and a shorter timescale of $\log (\tau_j/\mathrm{day}) = 0.74^{+0.34}_{-0.28}$, indicating faster variability compared to the disk component. These values fall within the expected prior bounds. The H$\beta$ emission line light curve was modeled by convolving the optical V-band continuum with a Gaussian transfer function. The amplitude $\log f_{H\beta}$ was determined to be $-0.12^{+0.08}_{-0.09}$, and the mean time delay $\tau_{H\beta} = 84.8^{+23.2}_{-18.4}$ days, consistent with the uniform prior (0, 300) days as adopted in P2022. The slightly larger H$\beta$ lag, compared to previous measurements, is attributed to the removal of jet contributions. The width of the transfer function was $\log (\omega_{H\beta}/\mathrm{day}) = 1.57^{+0.36}_{-0.40}$. To determine the time delay between the radio emission and the optical jet continuum, we modeled the 37 GHz radio light curve using a Gaussian transfer function. The amplitude was $\log f_j = 0.77^{+0.52}_{-0.20}$ with the mean time delay $\tau_j = 41.6^{+18.5}_{-11.5}$ days, this was much smaller than the H$\beta$ BLR lag. The computed width of the transfer function was $\log (\omega_j/\mathrm{day}) = 1.50^{+0.07}_{-0.11}$.

\subsection{Jet–Disk Decomposition Using Polarization}
The key parameter defining the ratio of the total optical disk emission to the total optical emission, $f_d$, is allowed to vary within the range (0, 1). For PKS1510, the disk-to-total flux ratio is $f_d = 0.37^{+0.03}_{-0.03}$. This suggests a relatively stronger jet contribution compared to the disk in the total optical emission. The corresponding inferred polarization parameters are $\bar{p_j} = 3.60^{+0.03}_{-0.03}$\% and $\theta_j = 78.8^{+0.1}_{-0.1}\,^{\circ}$. In contrast, for PKS0736, the inferred disk contribution ratio is $f_d = 0.70^{+0.01}_{-0.01}$, indicating a weaker jet contribution, so the disk dominates the total optical emission. The optical jet average polarization degree is constrained to $\bar{p_j} = 2.05^{+0.03}_{-0.02}$\%, and the polarization angle to $\theta_j = 91.8^{+0.5}_{-0.5}\,^{\circ}$. 

The reconstructed light curves for PKS1510 and PKS0736 are presented in Fig.~\ref{fig:LC_recon}. The first panel shows the total optical continuum (black), along with the decomposed jet (goldenrod) and disk (blueviolet) components as functions of time in JD $-\,2,400,000$. The second panel displays the reconstructed H$\beta$ emission line light curve over the observed data, and the third panel illustrates the modeled and observed 37 GHz radio light curves. For PKS0736, both disk and jet components exhibit variability, with the jet fraction consistently lower than that of the disk. However, for PKS1510, the jet fraction appears high relative to the disk emission. 

\section{Discussion}
\label{sec:discussion}
We modeled the V-band continuum, H$\beta$ emission line, and 37 GHz Radio light curves assuming certain priors. The Bayesian framework reveals the posterior distribution for the parameters, and this is executed by the code \texttt{DECOMP} developed by Li2020. 
\subsection{Previous Studies}
\textit{PKS1510} has been investigated through long-term spectroscopic monitoring as presented in R2020. The source displays strong optical variability, with fractional rms variations of $37.30 \pm 0.06$\% in the continuum ($f_{5100}$) and $11.88 \pm 0.29$\% in the H$\beta$ emission line, indicating significant contributions from both the accretion disk and the broad-line region. Cross-correlation analysis yielded a observed-frame H$\beta$ lag of $61.1^{+4.40}_{-3.20}$ days, corresponding to a BLR size of $44.9^{+2.9}_{-2.3}$ light-days(rest frame), which supports a thermal disk origin. However, jet contributions are evident from multi-wavelength observational signatures. The analysis of the non-thermal dominance (NTD) parameter revealed that for $\log L_{5100} \gtrsim 45.6$ erg\,s$^{-1}$, NTD values exceeded 2, suggesting enhanced non-thermal contamination in the optical continuum during high-flux states (beyond MJD 57100). The correlation between $L_{\mathrm{H\beta}}$ and $L_{5100}$ for the full sample was moderate ($r_s = 0.39$, $p = 10^{-11}$), but improved substantially for epochs with NTD $<$ 2 ($r_s = 0.71$, $p = 10^{-35}$), indicating that the BLR reprocessing signal is more apparent when the jet is in a quiescent state. Consistent with this, decade-long spectroscopy shows a BLR “breathing” anti-correlation between H$\beta$ width and line luminosity and contemporaneous H$\beta$ brightening during optical jet flares, implying that the jet can ionize BLR clouds and that its base lies within the BLR; using spectra in disk-dominated epochs yields a revised black hole mass of $M_{\rm BH}= (2.85\pm0.37)\times10^{8}\,M_\odot$ with minimal jet contamination \citep{amador2025}. 
Multi-wavelength timing further indicates near-zero optical/NIR–$\gamma$ lags (co-spatial synchrotron$+$SSC) but $\sim$60\,day X-ray delays (external Compton), consistent with a stratified jet \citep{amador2024}. Broad-band SED fits place the $\gamma$-ray zone just beyond the BLR yet within the dusty torus, favor external Compton on BLR photons, and suggest state-dependent inner-disk structure, with a mildly magnetized, matter-dominated outflow likely powered via Blandford–Znajek \citep{lei2025}. Independent spectropolarimetry shows high, variable continuum polarization but essentially unpolarized broad lines, reinforcing a synchrotron-dominated continuum atop a quasar-like BLR \citep{podjed2024}. 

For PKS1510, our finding of a 37\% disk contribution is slightly lower with \citet{Deng2026}, who found the jet and disk components contribute comparably. However, our results show a slightly higher jet fraction for this source. In the case of PKS0736, although the source is categorized as primarily jet-dominated with an average $\mathrm{NTD}$ of 2.7, our 70\% disk fraction represents a successful isolation of the thermal continuum during quiescent jet states. This supports the paper's conclusion that disk fluxes can be effectively recovered in 72\% of epochs through specialized decomposition, even when the jet is the long-term dominant feature.

In the present work, \texttt{DECOMP} modeling facilitated the separation of disk and jet components in PKS1510. The disk variability was characterized by $\log \sigma_d = -1.31^{+0.05}_{-0.06}$ and a short damping timescale of $\log (T_d/\mathrm{day}) = 1.26^{+0.12}_{-0.11}$, indicative of high-cadence variations. In contrast, the jet component showed a lower variability amplitude and a longer variability timescale. The H$\beta$ time delay derived using the decomposed disk continuum was $\tau_{H\beta} = 54.4^{+4.4}_{-5.4}$ days, slightly shorter than the previously reported value. The disk-to-total flux ratio $f_d = 0.37^{+0.03}_{-0.03}$ suggests that while the jet emission dominates, the disk still contributes to the optical flux. The decomposed disk component exhibits higher flux variability than the isolated jet emission.

For \textit{PKS0736}, a detailed RM analysis was conducted by P2022. The source exhibited moderate variability in emission lines, with fractional rms values of 21.10\% for H$\beta$ and 29.73\% for H$\gamma$, with higher variability in the V-band of 69\%. During the flaring phase between MJD 57000–57285, significant $\gamma$-ray and optical continuum variability was detected, indicating a strong jet contribution. To quantify this, the non-thermal dominance (NTD) parameter was examined. Assuming the H$\beta$ emission is driven by thermal optical variations originating from the accretion disk, the NTD value exceeded 2 for $\log L_{5100} > 44.6$\,erg\,s$^{-1}$, suggesting a strong jet component at higher luminosities. The correlation between $L_{\mathrm{H\beta}}$ and $L_{5100}$ displayed increased scatter beyond this threshold. A stronger Spearman correlation was observed for epochs with NTD $<$ 2 ($r_s = 0.71$, $p = 10^{-7}$) compared to the full sample ($r_s = 0.61$, $p = 10^{-15}$), affirming that the emission line variability is more closely tied to the disk when the jet contribution is minimal. The \texttt{DECOMP} framework decomposes the total optical emission into disk and jet components. The optical V-band continuum was modeled as a Damped Random Walk (DRW) process, and the H$\beta$ emission line light curve was reconstructed by convolving the modeled disk continuum with a Gaussian transfer function, yielding a time delay of $\tau_{H\beta} = 84.8^{+23.2}_{-18.4}$ days. This is comparatively slightly longer than the previously reported value of $\sim79$ days (observed frame), likely due to the removal of jet contamination, resulting in a cleaner disk signal that more accurately traces the BLR response. The derived disk contribution ratio $f_d = 0.70^{+0.01}_{-0.01}$ indicates that approximately 30\% of the optical continuum arises from the jet. The inferred average polarization degree $\bar{p_j} = 2.05^{+0.03}_{-0.02}$\% and angle $\theta_j = 91.8^{+0.5}_{-0.5}$\,$^{\circ}$ further support the presence of a structured jet component.

\subsection{Methodological Limitations of the Light Curve Decomposition}
\label{sec:decomp_limitations}
While our decomposition separates the thermal and non-thermal contributions to the optical continuum, the method has limitations for highly variable blazars. Following \citet{Li2020}, the optical jet component is modelled from a radio light curve, here the 37~GHz light curve, and can therefore only reproduce variations on the timescales present in the radio band. We adopt 37~GHz rather than the 15~GHz used by \citet{Li2020} for 3C~273, as higher-frequency radio emission arises closer to the optical emission zone. Nevertheless, the 37~GHz emission originates in a synchrotron self-absorbed region located downstream of the optical emission zone, and its variations are slower than, and delayed with respect to, those in the optical. For PKS1510, \citet{Yuan2023} found simultaneous optical and $\gamma$-ray variations followed by a delayed 37~GHz response. Optical jet variability on shorter timescales, generally attributed to a more compact emission region upstream of the 37~GHz-emitting zone, is thus not captured by our jet template.

The effect of this limitation is evident in Fig.~\ref{fig:LC_recon} for PKS1510. The modelled jet component shows low variability, whereas the disk component shows short, high-amplitude spikes coinciding with flares in the total flux. Such rapid, large-amplitude variations are more characteristic of jet emission than of thermal disk variability. These spikes are not reflected in the H$\beta$ light curve, which varies smoothly, although short-timescale variations would in any case be strongly smoothed by the BLR response over its light-crossing time. The disk component of PKS1510 therefore likely contains residual jet emission on short timescales, and the derived disk fraction and H$\beta$ lag for this source should be interpreted with caution. A more complete separation would require a multi-zone description of the jet emission, which is beyond the scope of this work.
\subsection{Radius Luminosity relation}
\begin{figure*}
    \centering
    \includegraphics[width=10cm, height=9cm]{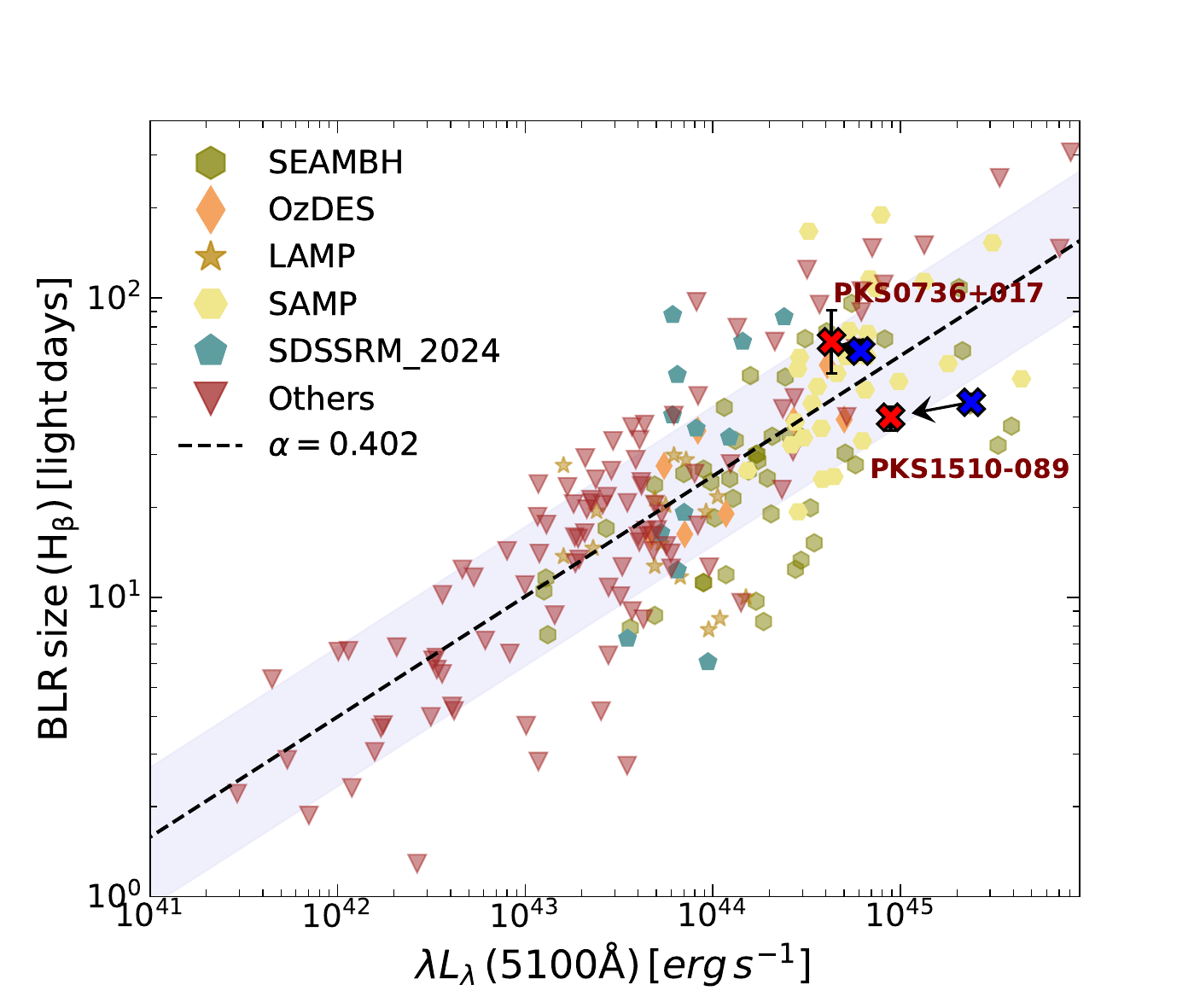}
    \caption{The plot presents the H$\beta$ broad-line region (BLR) size and the optical continuum luminosity at $5100\,\text{\AA}$ relation. A variety of reverberation-mapped (RM) AGN samples from the literature are overplotted for comparison. These include SEAMBHs \citep[][grey open circles]{Du2016a, Du2018, Hu2021, Li2021}, SDSSRM-2024 sources \citep[][grey open squares]{Shen2024}, OzDES AGNs \citep[][grey open diamonds]{Malik2023}, LAMP sources \citep[][grey open circles]{U2022}, and SAMP sources \citep[][grey open hexagons]{Woo2024}. Additional RM AGNs from various studies \citep[][grey open inverted triangles]{Bentz2013, Park_2017, Rakshit2019, Bonta2020, Rakshit2020, Pandey2022} are also included for completeness. The black dashed line indicates the best-fit $R$–$L$ relation as reported by \citet{Woo2024} with its intrinsic scatter ($\pm$0.23 dex, shown as a shaded lavender band) are overplotted. Our \texttt{DECOMP} target sources, PKS~0736+017 and PKS~1510-089, are highlighted with filled blue stars, representing their previously reported H$\beta$ BLR sizes and continuum luminosities. The updated measurements obtained using \texttt{DECOMP}, which isolates the disk-only optical continuum, are shown in red. Notably, PKS~1510-089 shifts much closer to the best-fit $R$–$L$ relation, indicating reduced deviation compared to earlier estimates. In contrast, PKS~0736+017 moves a bit further away from the relation, suggesting an increased discrepancy.}
    \label{fig:R_L}
\end{figure*}
For PKS1510, the mean observed monochromatic luminosity at $5100\,\text{\AA}$ was reported as $L_{5100}=2.38\times10^{45}\,\mathrm{erg\,s^{-1}}$ by R2020. To gauge the thermal (disk) fraction within this observed continuum, they used the non-thermal dominance (NTD) metric and, during quiescent epochs, obtained a median $\mathrm{NTD}=1.67^{+0.31}_{-0.23}$, implying a disk contribution of $\simeq60\%$ to the optical continuum. Independent broad-band SED modelling across active and quiescent states in 2015 further supports substantial thermal emission: \citet{Prince2019} inferred a $5100\,\text{\AA}$ disk fraction of $\approx44\%$ in the quiescent window MJD~$57108$–$57208$ using a time-independent two-zone emission model. In our decomposition with \texttt{DECOMP}, we find a somewhat smaller disk fraction, $f_d=0.37\pm0.01$, indicating that $\sim37\%$ of the optical flux arises from the accretion disk. Adopting this value yields a revised disk-only luminosity of $L_{5100}^{\rm disk}\!=\!f_d\,L_{5100}\!\approx\!8.85\times10^{44}\,\mathrm{erg\,s^{-1}}$, smaller than that of the NTD-based, and consistent with SED-based estimates. This underscores the importance of explicitly correcting for jet contamination in reverberation analyses. Consequently, the source’s locus in the $R_{\rm BLR}$–$L_{5100}$ plane (Fig.~\ref{fig:R_L}) shifts lower leftward to the best-fit $R_{\rm BLR}$–$L_{5100}^{\rm disk}$ relation.

For PKS0736, P2022 reported a mean continuum luminosity at $5100\,\text{\AA}$ of $L_{5100}^{\mathrm{obs}} = 6.13 \times 10^{44}$\,erg\,s$^{-1}$. To separate the thermal and non-thermal components, they estimated an NTD of 2.51 during a quiescent phase, implying a disk contribution of only $\sim$33\%. From this, they derived a disk-only luminosity of $L_{5100}^{\mathrm{disk}} = 1.44 \times 10^{44}$\,erg\,s$^{-1}$. Previous studies have estimated the bolometric luminosity of PKS0736 to range between $10^{44.6}$ and $10^{45.7}$\,erg\,s$^{-1}$ \citep{McLure2001, Dai2007, Abdalla2020}. Although no previous estimate of the disk fraction at $5100\,\text{\AA}$ has been reported for this source, the spectral decomposition shown in Fig.~5 of \citet{Abdalla2020} allows one to infer that the accretion disk contributes roughly $50$–$60\%$ of the optical emission, with the remainder arising from the jet. In this work, our decomposition using \texttt{DECOMP} inferred a disk contribution fraction of $\approx$70\%. Applying this to the observed luminosity yields a revised disk-only luminosity of $4.30 \times 10^{44}$\,erg\,s$^{-1}$. This updated value and the corresponding revised BLR size are shown in the $R_{\mathrm{BLR}}$–$L_{5100}$ plot in Fig.~\ref{fig:R_L}, where the source shifts towards the upper left and now exhibits a comparatively larger deviation but still within the uncertainity limit from the fitted $R$–$L$ relation compared to earlier estimates.

\subsubsection{Deviation in R–L relation from FeII contribution}
Previous studies of high accretion rate, strong Fe \textsc{ii} emitters have shown significant departures from the canonical $R_{\rm BLR}$–$L_{5100}$ relation \citep[e.g.,][]{Du2016a,Du2018}. In the slim disk framework, pronounced self-shadowing can imprint a highly anisotropic ionising field, partitioning the BLR into two dynamically distinct zones and naturally biasing reverberation lags \citep{Wang2014}. Building on this, \citet{Du2019} argued that the Fe\textsc{ii} strength ($R_{\mathrm{FeII}}=F_{\mathrm{FeII}}/F_{\mathrm{H}\beta}$) is a principal driver of shortened lags in high accreting AGNs and provided a scaling that explicitly includes $R_{\mathrm{FeII}}$. 

For PKS1510 ($L_{\rm disk}=8.85\times10^{44}\,{\rm erg\,s^{-1}}$), the rest-frame BLR size of $\sim 40$ light-days is a factor of $\sim 2.7$ shorter than the \citet{Bentz2013} prediction ($\sim 108$ light-days) and a factor of $\sim 1.5$ smaller than the \citet{Woo2024} estimate ($\sim 61$ light-days). Inclusion of $R_{\mathrm{FeII}}=0.52\pm0.09$ in the \citet{Du2019} multivariate relation yields a prediction of $\sim 78$ light-days. While this reduces the expected radius, it remains significantly longer than the measured BLR size, which was obtained after correcting for jet contamination. PKS1510, therefore, sits well below both the \citet{Bentz2013} and \citet{Woo2024} best-fit relations, remaining consistent with the interpretation that high-Eddington sources can exhibit shortened BLR sizes owing to slim-disk anisotropy and self-shadowing.

For PKS0736 ($L_{\rm disk}=4.30\times10^{44}\,{\rm erg\,s^{-1}}$), the observed BLR size of $\sim 71.3$ light-days is remarkably consistent with the \citet{Bentz2013} prediction ($\sim 73$ light-days), though it remains longer than the \citet{Woo2024} estimate ($\sim 46$ light-days). The \citet{Du2019} prediction, including $R_{\rm FeII}=0.58$, yields an expected radius of $\sim 54$ light-days. Although the moderately strong $R_{\rm FeII}$ theoretically shifts the expected BLR size to smaller values, the measured lag actually adheres closely to the classical $R$-$L$ relation. 

Together, these two quasars illustrate opposite departures from the $R$–$L$ relation: PKS1510 lies well below, consistent with high-Eddington induced shortening, while PKS0736 lies well above, showing an unusually extended BLR size. This contrast highlights that while $R_{\rm FeII}$ can partially account for lag shortening in some sources, it does not universally reconcile deviations from the $R$–$L$ relation.

Additionally, the comparison of BLR size predictions between \citet{Bentz2013} and \citet{Woo2024} reveals systematic differences. The \citet{Bentz2013} relation has a steeper slope and a higher intercept, resulting in larger predicted BLR sizes at fixed luminosity. In contrast, \citet{Woo2024} found a shallower slope ($\alpha \approx 0.40$) and lower intercept ($K \approx 1.40$), which systematically reduce predicted BLR sizes, consistent with their interpretation that a higher fraction of high-Eddington ratio AGNs in luminous bins drives the shallower relation. \citet{Woo2024} further argued that the BLR can become shortened in high-accreting systems owing to slim-disk self-shadowing, inclination effects, and the partitioning of the BLR structure. Their analysis of Eddington ratio distributions (Fig.~21 of \citet{Woo2024}) shows that higher-luminosity bins contain higher average $\lambda_{\rm Edd}$, naturally leading the $R$–$L$ slope to smaller values.

\section{Conclusion}
\label{sec:conclusion}
We decomposed the optical light curve of two flat-spectrum radio quasars, PKS0736 and PKS1510, into thermal (disk) and non-thermal (jet) components. Toward this, we utilized multi-epoch optical photometric and polarimetric data from the SO blazar monitoring program along with ASAS-SN and SMARTS observations. Both sources display variable levels of optical polarization throughout the observing window. Additionally, we incorporated optical continuum and H$\beta$ emission line light curves for PKS0736 and PKS1510, along with 37 GHz radio light curves from the 13.7~m diameter Aalto University Mets\"{a}hovi radio telescope.

For PKS1510, the disk-to-total optical flux ratio was tightly constrained, with $f_{\mathrm{d}} = 0.37^{+0.01}_{-0.01}$, indicating that the jet dominated the observed optical continuum. Using the decomposed disk light curve, we obtained a revised H$\beta$ lag of $54.4^{+4.4}_{-5.4}$ days. The 37~GHz emission lagged the optical jet component by $8.3^{+2.5}_{-1.4}$ days,  consistent with the delayed radio response relative to the optical reported for this source by \citet{Yuan2023}. Because the modelled jet component followed the more slowly varying 37~GHz light curve, it captured the long-term jet contribution but not the rapid optical jet flares. As a result, the jet component showed comparatively low variability, while the disk component showed short, high-amplitude spikes coinciding with flares in the total flux. The disk component of PKS1510 therefore likely contained residual short-timescale jet emission, and its disk fraction and H$\beta$ lag should be interpreted with caution.

The disk luminosity estimated from our decomposition was lower than earlier values based on non-thermal dominance corrections, which underscores the importance of incorporating polarization information in decomposition analyses. After correcting for the jet contribution, PKS1510 lay much closer to the best-fit $R_{\mathrm{BLR}}$-$L_{5100}$ relation, although its position may also have been affected by its strong $R_{\mathrm{Fe\,II}}$ and by the uncertainty in the disk fraction.

For PKS0736, the polarization modeling constructs a scenario in which the disk dominates the total optical emission, with a decomposed disk fraction of 70\%, which aligns well with the previous estimate of NTD-based corrections. The H$\beta$ lag derived using the jet-subtracted disk continuum is longer at $84.8^ {+23.2}_{-18.4}$ days, than the value reported in earlier work of 78 days using the total continuum. After correcting the jet contribution in the optical emission, PKS0736 lies a bit above the best fit $R_{\mathrm{BLR}}$–$L_{5100}$. The radio jet is delayed by $41.6^{+18.5}_{-11.5}$ days with respect to the continuum. 

\section*{Acknowledgements}
S.~P. acknowledges support from the European Research Council (ERC) under the European Union’s Horizon 2020 research and innovation program (grant No.~951549). 
S.~P. expresses gratitude to Yan-Rong Li for significant and informative discussions regarding the software \texttt{DECOMP}; his insights were instrumental in the modification and interpretation of the source codes. S.~P. also thanks Bożena Czerny for helpful discussions regarding the simulation results. This publication makes use of data obtained at the Mets\"{a}hovi Radio Observatory, operated by Aalto University in Finland. The contributions of the various diligent observers at Aalto University are gratefully acknowledged.

\section*{Data Availability}
The data underlying this paper were obtained from the Steward Observatory database via \url{http://james.as.arizona.edu/~psmith/Fermi/}, and all derived data generated during this research are provided in the article and its online supplementary material. This work also makes use of up-to-date SMARTS optical/near-infrared light curves available at \url{www.astro.yale.edu/smarts/glast/home.php}. Furthermore, we incorporate data from the All-Sky Automated Survey for Super-Novae (ASAS-SN) and the Metsähovi Radio Observatory, which is operated by Aalto University in Finland.

\facilities{Steward Observatory (SO), Small and Moderate Aperture Research Telescope System(SMARTS), 13.7~m diameter Aalto University Mets\"{a}hovi radio telescope, All-Sky Automated Survey for Supernovae (ASAS-SN)}
\software{\texttt{DECOMP} \citep{Li2020}, CDNest \citep{Li2020zndo...3884449L}, corner.py \citep{Foreman-Mackey2016}}, {\tt PyCALI} \citep{Pycali2014}


\begin{thebibliography}{}
\expandafter\ifx\csname natexlab\endcsname\relax\def\natexlab#1{#1}\fi
\providecommand{\url}[1]{\href{#1}{#1}}
\providecommand{\dodoi}[1]{doi:~\href{http://doi.org/#1}{\nolinkurl{#1}}}
\providecommand{\doeprint}[1]{\href{http://ascl.net/#1}{\nolinkurl{http://ascl.net/#1}}}
\providecommand{\doarXiv}[1]{\href{https://arxiv.org/abs/#1}{\nolinkurl{https://arxiv.org/abs/#1}}}

\bibitem[{A. {Amador-Portes} {et~al.}(2025){Amador-Portes}, {Chavushyan}, {Pati{\~n}o-{\'A}lvarez}, \& {Ram{\'o}n-Vald{\'e}s}}]{amador2025}
{Amador-Portes}, A., {Chavushyan}, V., {Pati{\~n}o-{\'A}lvarez}, V.~M., \& {Ram{\'o}n-Vald{\'e}s}, J. 2025, \bibinfo{title}{{Unveiling the Emission Mechanisms of Blazar PKS 1510{\ensuremath{-}}089. II. Jet{\ensuremath{-}}BLR Connection and Black Hole Mass Estimation},} \apj, 979, 227, \dodoi{10.3847/1538-4357/ada38b}

\bibitem[{A. {Amador-Portes} {et~al.}(2024){Amador-Portes}, {Garc{\'\i}a-P{\'e}rez}, {Chavushyan}, \& {Pati{\~n}o-{\'A}lvarez}}]{amador2024}
{Amador-Portes}, A., {Garc{\'\i}a-P{\'e}rez}, A., {Chavushyan}, V., \& {Pati{\~n}o-{\'A}lvarez}, V.~M. 2024, \bibinfo{title}{{Unveiling the Emission Mechanisms of Blazar PKS 1510-089. I. Multiwavelength Variability},} \apj, 977, 178, \dodoi{10.3847/1538-4357/ad8ddd}

\bibitem[{J.~N. {Bahcall} {et~al.}(1972){Bahcall}, {Kozlovsky}, \& {Salpeter}}]{Bahcall1972}
{Bahcall}, J.~N., {Kozlovsky}, B.-Z., \& {Salpeter}, E.~E. 1972, \bibinfo{title}{{On the Time Dependence of Emission-Line Strengths from a Photoionized Nebula},} \apj, 171, 467, \dodoi{10.1086/151300}

\bibitem[{M.~C. Bentz {et~al.}(2013)Bentz, Denney, Grier, Barth, Peterson, Vestergaard, Bennert, Canalizo, De~Rosa, Filippenko, Gates, Greene, Li, Malkan, Pogge, Stern, Treu, \& Woo}]{Bentz2013}
Bentz, M.~C., Denney, K.~D., Grier, C.~J., {et~al.} 2013, \bibinfo{title}{The {Low}-luminosity {End} of the {Radius}-{Luminosity} {Relationship} for {Active} {Galactic} {Nuclei},} ApJ, 767, 149, \dodoi{10.1088/0004-637X/767/2/149}

\bibitem[{R.~D. {Blandford} \& C.~F. {McKee}(1982){Blandford} \& {McKee}}]{Blandford1982}
{Blandford}, R.~D., \& {McKee}, C.~F. 1982, \bibinfo{title}{{Reverberation mapping of the emission line regions of Seyfert galaxies and quasars.},} \apj, 255, 419, \dodoi{10.1086/159843}

\bibitem[{E. {Bonning} {et~al.}(2012){Bonning}, {Urry}, {Bailyn}, {Buxton}, {Chatterjee}, {Coppi}, {Fossati}, {Isler}, \& {Maraschi}}]{Bonning2012}
{Bonning}, E., {Urry}, C.~M., {Bailyn}, C., {et~al.} 2012, \bibinfo{title}{{SMARTS Optical and Infrared Monitoring of 12 Gamma-Ray Bright Blazars},} \apj, 756, 13, \dodoi{10.1088/0004-637X/756/1/13}

\bibitem[{E.~D. Bont{\`{a}} {et~al.}(2020)Bont{\`{a}}, Peterson, Bentz, Brandt, Ciroi, Rosa, Alvarez, Grier, Hall, {Hern{\'{a}}ndez Santisteban}, Ho, Homayouni, Horne, Kochanek, Li, Morelli, Pizzella, Pogge, Schneider, Shen, Trump, \& Vestergaard}]{Bonta2020}
Bont{\`{a}}, E.~D., Peterson, B.~M., Bentz, M.~C., {et~al.} 2020, \bibinfo{title}{{The Sloan Digital Sky Survey Reverberation Mapping Project: Estimating Masses of Black Holes in Quasars with Single-epoch Spectroscopy},} ApJ, 903, 112, \dodoi{10.3847/1538-4357/ABBC1C}

\bibitem[{B.~J. {Brewer} {et~al.}(2011){Brewer}, {P{\'a}rtay}, \& {Cs{\'a}nyi}}]{Brewer2011}
{Brewer}, B.~J., {P{\'a}rtay}, L.~B., \& {Cs{\'a}nyi}, G. 2011, \bibinfo{title}{{Diffusive nested sampling},} Statistics and Computing, 21, 649, \dodoi{10.1007/s11222-010-9198-8}

\bibitem[{H. {Dai} {et~al.}(2007){Dai}, {Xie}, {Zhou}, {Li}, {Chen}, \& {Ma}}]{Dai2007}
{Dai}, H., {Xie}, G.~Z., {Zhou}, S.~B., {et~al.} 2007, \bibinfo{title}{{Correlation between Eddington Ratios and Broad-Line Luminosities in Flat-Spectrum Radio Quasars, BL Lacertae Objects, and Radio Galaxies},} \aj, 133, 2187, \dodoi{10.1086/511769}

\bibitem[{J. {Deng} {et~al.}(2026){Deng}, {Liu}, {Wang}, \& {Jiang}}]{Deng2026}
{Deng}, J., {Liu}, L., {Wang}, Y., \& {Jiang}, Y. 2026, \bibinfo{title}{{A Novel Reverberation Mapping Method for Blazars},} \apj, 997, 258, \dodoi{10.3847/1538-4357/ae260a}

\bibitem[{P. Du \& J.-M. Wang(2019)Du \& Wang}]{Du2019}
Du, P., \& Wang, J.-M. 2019, \bibinfo{title}{{The Radius–Luminosity Relationship Depends on Optical Spectra in Active Galactic Nuclei},} The Astrophysical Journal, 886, 42, \dodoi{10.3847/1538-4357/ab4908}

\bibitem[{P. Du {et~al.}(2016)Du, Lu, Zhang, Huang, Wang, Hu, Qiu, Li, Fan, Fang, Bai, Bian, Yuan, Ho, Wang, \& collaboration)}]{Du2016a}
Du, P., Lu, K.-X., Zhang, Z.-X., {et~al.} 2016, \bibinfo{title}{SUPERMASSIVE BLACK HOLES WITH HIGH ACCRETION RATES IN ACTIVE GALACTIC NUCLEI. V. A NEW SIZE–LUMINOSITY SCALING RELATION FOR THE BROAD-LINE REGION,} The Astrophysical Journal, 825, 126, \dodoi{10.3847/0004-637X/825/2/126}

\bibitem[{P. Du {et~al.}(2018)Du, Zhang, Wang, Huang, Zhang, Lu, Hu, Li, Bai, Bian, Yuan, Ho, Wang, \& collaboration)}]{Du2018}
Du, P., Zhang, Z.-X., Wang, K., {et~al.} 2018, \bibinfo{title}{Supermassive Black Holes with High Accretion Rates in Active Galactic Nuclei. IX. 10 New Observations of Reverberation Mapping and Shortened Hβ Lags,} The Astrophysical Journal, 856, 6, \dodoi{10.3847/1538-4357/aaae6b}

\bibitem[{E.~L. {Fitzpatrick}(1999){Fitzpatrick}}]{Fitzpatrick1999}
{Fitzpatrick}, E.~L. 1999, \bibinfo{title}{{Correcting for the Effects of Interstellar Extinction},} \pasp, 111, 63, \dodoi{10.1086/316293}

\bibitem[{D. {Foreman-Mackey}(2016){Foreman-Mackey}}]{Foreman-Mackey2016}
{Foreman-Mackey}, D. 2016, \bibinfo{title}{{corner.py: Scatterplot matrices in Python},} The Journal of Open Source Software, 1, 24, \dodoi{10.21105/joss.00024}

\bibitem[{K. {Hart} {et~al.}(2023){Hart}, {Shappee}, {Hey}, {Kochanek}, {Stanek}, {Lim}, {Dobbs}, {Tucker}, {Jayasinghe}, {Beacom}, {Boright}, {Holoien}, {Ong}, {Prieto}, {Thompson}, \& {Will}}]{Hart2023}
{Hart}, K., {Shappee}, B.~J., {Hey}, D., {et~al.} 2023, \bibinfo{title}{{ASAS-SN Sky Patrol V2.0},} arXiv e-prints, arXiv:2304.03791, \dodoi{10.48550/arXiv.2304.03791}

\bibitem[{C. Hazard {et~al.}(2018)Hazard, Jauncey, Goss, \& Herald}]{Hazard2018}
Hazard, C., Jauncey, D., Goss, W.~M., \& Herald, D. 2018, \bibinfo{title}{The Sequence of Events that led to the 1963 Publications in Nature of 3C 273, the First Quasar and the First Extragalactic Radio Jet,} Publications of the Astronomical Society of Australia, 35, e006, \dodoi{10.1017/pasa.2017.62}

\bibitem[{C. {Heiles}(2000){Heiles}}]{Heiles2000}
{Heiles}, C. 2000, \bibinfo{title}{{9286 Stars: An Agglomeration of Stellar Polarization Catalogs},} \aj, 119, 923, \dodoi{10.1086/301236}

\bibitem[{ {H.E.S.S. Collaboration} {et~al.}(2020){H.E.S.S. Collaboration}, {Abdalla}, {Adam}, {Aharonian}, {Ait Benkhali}, {Ang{\"u}ner}, {Arakawa}, {Arcaro}, {Armand}, {Ashkar}, \& et~al.}]{Abdalla2020}
{H.E.S.S. Collaboration}, {Abdalla}, H., {Adam}, R., {et~al.} 2020, \bibinfo{title}{{H.E.S.S. detection of very high-energy {\ensuremath{\gamma}}-ray emission from the quasar PKS 0736+017},} \aap, 633, A162, \dodoi{10.1051/0004-6361/201935906}

\bibitem[{C. {Hu} {et~al.}(2021){Hu}, {Li}, {Yang}, {Yang}, {Guo}, {Bao}, {Jiang}, {Du}, {Li}, {Xiao}, \& et~al.}]{Hu2021}
{Hu}, C., {Li}, S.-S., {Yang}, S., {et~al.} 2021, \bibinfo{title}{{Supermassive Black Holes with High Accretion Rates in Active Galactic Nuclei. XII. Reverberation Mapping Results for 15 PG Quasars from a Long-duration High-cadence Campaign},} \apjs, 253, 20, \dodoi{10.3847/1538-4365/abd774}

\bibitem[{C.~S. {Kochanek} {et~al.}(2017){Kochanek}, {Shappee}, {Stanek}, {Holoien}, {Thompson}, {Prieto}, {Dong}, {Shields}, {Will}, {Britt}, {Perzanowski}, \& {Pojma{\'n}ski}}]{Kochanek2017}
{Kochanek}, C.~S., {Shappee}, B.~J., {Stanek}, K.~Z., {et~al.} 2017, \bibinfo{title}{{The All-Sky Automated Survey for Supernovae (ASAS-SN) Light Curve Server v1.0},} \pasp, 129, 104502, \dodoi{10.1088/1538-3873/aa80d9}

\bibitem[{M. {Lei} {et~al.}(2025){Lei}, {Zheng}, {Zhang}, {Yuan}, \& {Wang}}]{lei2025}
{Lei}, M., {Zheng}, Y., {Zhang}, J., {Yuan}, Y., \& {Wang}, J. 2025, \bibinfo{title}{{Insight into the Origin of Multiwavelength Emissions of PKS 1510{\ensuremath{-}}089 Through Modeling 12 SEDs from 2008 to 2015},} \apj, 988, 13, \dodoi{10.3847/1538-4357/addf3d}

\bibitem[{S.-S. Li {et~al.}(2021)Li, Yang, Yang, Chen, Songsheng, Liu, Du, Luo, Yu, Hu, Jiang, Bao, Guo, Zhang, Li, Xiao, Lu, Ho, Bai, Bian, Aceituno, Minezaki, Horne, Kokubo, \& Wang}]{Li2021}
Li, S.-S., Yang, S., Yang, Z.-X., {et~al.} 2021, \bibinfo{title}{Reverberation Mapping of Two Luminous Quasars: The Broad-line Region Structure and Black Hole Mass,} The Astrophysical Journal, 920, 9, \dodoi{10.3847/1538-4357/ac116e}

\bibitem[{Y.-R. {Li}(2020){Li}}]{Li2020zndo...3884449L}
{Li}, Y.-R. 2020, \bibinfo{title}{{LiyrAstroph/CDNest: CDNest: A diffusive nested sampling code in C},}, v0.2.0 Zenodo, \dodoi{10.5281/zenodo.3884449}

\bibitem[{Y.-R. {Li} {et~al.}(2014){Li}, {Wang}, {Hu}, {Du}, \& {Bai}}]{Pycali2014}
{Li}, Y.-R., {Wang}, J.-M., {Hu}, C., {Du}, P., \& {Bai}, J.-M. 2014, \bibinfo{title}{{A Bayesian Method for the Intercalibration of Spectra in Reverberation Mapping},} \apjl, 786, L6, \dodoi{10.1088/2041-8205/786/1/L6}

\bibitem[{Y.-R. Li {et~al.}(2020)Li, Zhang, Jin, Du, Cui, Liu, \& Wang}]{Li2020}
Li, Y.-R., Zhang, Z.-X., Jin, C., {et~al.} 2020, \bibinfo{title}{Untangling Optical Emissions of the Jet and Accretion Disk in the Flat-spectrum Radio Quasar 3C 273 with Reverberation Mapping Data,} The Astrophysical Journal, 897, 18, \dodoi{10.3847/1538-4357/ab95a3}

\bibitem[{U. {Malik} {et~al.}(2023){Malik}, {Sharp}, {Penton}, {Yu}, {Martini}, {Lidman}, {Tucker}, {Davis}, {Lewis}, {Aguena}, {Allam}, {Alves}, {Andrade-Oliveira}, {Asorey}, {Bacon}, {Bertin}, {Bocquet}, {Brooks}, {Burke}, {Carnero Rosell}, {Carollo}, {Carrasco Kind}, {Carretero}, {Costanzi}, {da Costa}, {Pereira}, {De Vicente}, {Desai}, {Diehl}, {Doel}, {Everett}, {Ferrero}, {Frieman}, {Garc{\'\i}a-Bellido}, {Gerdes}, {Gruen}, {Gruendl}, {Gschwend}, {Hinton}, {Hollowood}, {Honscheid}, {James}, {Kuehn}, {Marshall}, {Mena-Fern{\'a}ndez}, {Menanteau}, {Miquel}, {Ogando}, {Palmese}, {Paz-Chinch{\'o}n}, {Pieres}, {Plazas Malag{\'o}n}, {Raveri}, {Rodriguez-Monroy}, {Romer}, {Sanchez}, {Scarpine}, {Sevilla-Noarbe}, {Smith}, {Soares-Santos}, {Suchyta}, {Swanson}, {Tarle}, {Taylor}, {Tucker}, {Weaverdyck}, \& {Wilkinson}}]{Malik2023}
{Malik}, U., {Sharp}, R., {Penton}, A., {et~al.} 2023, \bibinfo{title}{{OzDES Reverberation Mapping Program: H{\ensuremath{\beta}} lags from the 6-yr survey},} \mnras, 520, 2009, \dodoi{10.1093/mnras/stad145}

\bibitem[{A.~P. {Marscher} \& W.~K. {Gear}(1985){Marscher} \& {Gear}}]{Marscher1985}
{Marscher}, A.~P., \& {Gear}, W.~K. 1985, \bibinfo{title}{{Models for high-frequency radio outbursts in extragalactic sources, with application to the early 1983 millimeter-to-infrared flare of 3C 273.},} \apj, 298, 114, \dodoi{10.1086/163592}

\bibitem[{R.~J. {McLure} \& J.~S. {Dunlop}(2001){McLure} \& {Dunlop}}]{McLure2001}
{McLure}, R.~J., \& {Dunlop}, J.~S. 2001, \bibinfo{title}{{The black hole masses of Seyfert galaxies and quasars},} \mnras, 327, 199, \dodoi{10.1046/j.1365-8711.2001.04709.x}

\bibitem[{S. Pandey {et~al.}(2022)Pandey, Rakshit, Woo, \& Stalin}]{Pandey2022}
Pandey, S., Rakshit, S., Woo, J.-H., \& Stalin, C.~S. 2022, \bibinfo{title}{Spectroscopic reverberation mapping of {Quasar} {PKS} 0736+017: {Broad}-{Line} {Region} and {Black}-hole {Mass},} MNRAS, 516, 2671, \dodoi{10.1093/mnras/stac2418}

\bibitem[{S. Park {et~al.}(2017)Park, Woo, Romero-Colmenero, Crawford, Park, Cho, Jeon, Choi, Barth, Pei, Hickox, Sung, \& Im}]{Park_2017}
Park, S., Woo, J.-H., Romero-Colmenero, E., {et~al.} 2017, \bibinfo{title}{Reverberation Mapping of PG 0934+013 with the Southern African Large Telescope,} The Astrophysical Journal, 847, 125, \dodoi{10.3847/1538-4357/aa88a3}

\bibitem[{B.~M. Peterson(1993)Peterson}]{Peterson_1993}
Peterson, B.~M. 1993, \bibinfo{title}{Reverberation mapping of active galactic nuclei,} PASP, 105, 247, \dodoi{10.1086/133140}

\bibitem[{S.~A. {Podjed} {et~al.}(2024){Podjed}, {Hickox}, {Isler}, {B{\"o}ttcher}, \& {Schutte}}]{podjed2024}
{Podjed}, S.~A., {Hickox}, R.~C., {Isler}, J.~C., {B{\"o}ttcher}, M., \& {Schutte}, H.~M. 2024, \bibinfo{title}{{Optical Spectropolarimetric Variability Properties in Blazars PKS 0637-75 and PKS 1510-089},} \apj, 968, 130, \dodoi{10.3847/1538-4357/ad4111}

\bibitem[{R. Prince {et~al.}(2019)Prince, Gupta, \& Nalewajko}]{Prince2019}
Prince, R., Gupta, N., \& Nalewajko, K. 2019, \bibinfo{title}{Two-zone Emission Modeling of PKS 1510-089 during the High State of 2015,} The Astrophysical Journal, 883, 137, \dodoi{10.3847/1538-4357/ab3afa}

\bibitem[{S. Rakshit(2020)Rakshit}]{Rakshit2020}
Rakshit, S. 2020, \bibinfo{title}{{Broad line region and black hole mass of PKS 1510-089 from spectroscopic reverberation mapping},} Astronomy and Astrophysics, 642, 1, \dodoi{10.1051/0004-6361/202038324}

\bibitem[{S. Rakshit {et~al.}(2019)Rakshit, Woo, Gallo, Hodges-Kluck, Shin, Jeon, Bae, Baldassare, Cho, Cho, Foord, Kang, Kang, Karouzos, Kim, Kim, Le, Park, Park, Son, Sung, Bennert, \& Malkan}]{Rakshit2019}
Rakshit, S., Woo, J.-H., Gallo, E., {et~al.} 2019, \bibinfo{title}{{The Seoul National University AGN Monitoring Project. II. BLR Size and Black Hole Mass of Two AGNs},} The Astrophysical Journal, 886, 93, \dodoi{10.3847/1538-4357/ab49fd}

\bibitem[{E.~F. {Schlafly} \& D.~P. {Finkbeiner}(2011){Schlafly} \& {Finkbeiner}}]{Schlafly2011}
{Schlafly}, E.~F., \& {Finkbeiner}, D.~P. 2011, \bibinfo{title}{{Measuring Reddening with Sloan Digital Sky Survey Stellar Spectra and Recalibrating SFD},} \apj, 737, 103, \dodoi{10.1088/0004-637X/737/2/103}

\bibitem[{D.~J. {Schlegel} {et~al.}(1998){Schlegel}, {Finkbeiner}, \& {Davis}}]{Schlegel1998}
{Schlegel}, D.~J., {Finkbeiner}, D.~P., \& {Davis}, M. 1998, \bibinfo{title}{{Maps of Dust Infrared Emission for Use in Estimation of Reddening and Cosmic Microwave Background Radiation Foregrounds},} \apj, 500, 525, \dodoi{10.1086/305772}

\bibitem[{G.~D. {Schmidt} {et~al.}(1992){Schmidt}, {Stockman}, \& {Smith}}]{1992ApJ...398L..57S}
{Schmidt}, G.~D., {Stockman}, H.~S., \& {Smith}, P.~S. 1992, \bibinfo{title}{{Discovery of a Sub-Megagauss Magnetic White Dwarf through Spectropolarimetry},} \apjl, 398, L57, \dodoi{10.1086/186576}

\bibitem[{K. {Serkowski} {et~al.}(1975){Serkowski}, {Mathewson}, \& {Ford}}]{Serkowski1975}
{Serkowski}, K., {Mathewson}, D.~S., \& {Ford}, V.~L. 1975, \bibinfo{title}{{Wavelength dependence of interstellar polarization and ratio of total to selective extinction.},} \apj, 196, 261, \dodoi{10.1086/153410}

\bibitem[{B.~J. {Shappee} {et~al.}(2014){Shappee}, {Prieto}, {Grupe}, {Kochanek}, {Stanek}, {De Rosa}, {Mathur}, {Zu}, {Peterson}, {Pogge}, {Komossa}, {Im}, {Jencson}, {Holoien}, {Basu}, {Beacom}, {Szczygie{\l}}, {Brimacombe}, {Adams}, {Campillay}, {Choi}, {Contreras}, {Dietrich}, {Dubberley}, {Elphick}, {Foale}, {Giustini}, {Gonzalez}, {Hawkins}, {Howell}, {Hsiao}, {Koss}, {Leighly}, {Morrell}, {Mudd}, {Mullins}, {Nugent}, {Parrent}, {Phillips}, {Pojmanski}, {Rosing}, {Ross}, {Sand}, {Terndrup}, {Valenti}, {Walker}, \& {Yoon}}]{Shappee2014}
{Shappee}, B.~J., {Prieto}, J.~L., {Grupe}, D., {et~al.} 2014, \bibinfo{title}{{The Man behind the Curtain: X-Rays Drive the UV through NIR Variability in the 2013 Active Galactic Nucleus Outburst in NGC 2617},} \apj, 788, 48, \dodoi{10.1088/0004-637X/788/1/48}

\bibitem[{M.~S. {Shaw} {et~al.}(2012){Shaw}, {Romani}, {Cotter}, {Healey}, {Michelson}, {Readhead}, {Richards}, {Max-Moerbeck}, {King}, \& {Potter}}]{Shaw2012}
{Shaw}, M.~S., {Romani}, R.~W., {Cotter}, G., {et~al.} 2012, \bibinfo{title}{{Spectroscopy of Broad-line Blazars from 1LAC},} \apj, 748, 49, \dodoi{10.1088/0004-637X/748/1/49}

\bibitem[{Y. {Shen} {et~al.}(2024){Shen}, {Grier}, {Horne}, {Stone}, {Li}, {Yang}, {Homayouni}, {Trump}, {Anderson}, {Brandt}, {Hall}, {Ho}, {Jiang}, {Petitjean}, {Schneider}, {Tao}, {Donnan}, {AlSayyad}, {Bershady}, {Blanton}, {Bizyaev}, {Bundy}, {Chen}, {Davis}, {Dawson}, {Fan}, {Greene}, {Gr{\"o}ller}, {Guo}, {Ibarra-Medel}, {Jiang}, {Keenan}, {Kollmeier}, {Lejoly}, {Li}, {de la Macorra}, {Moe}, {Nie}, {Rossi}, {Smith}, {Tee}, {Weijmans}, {Xu}, {Yue}, {Zhou}, {Zhou}, \& {Zou}}]{Shen2024}
{Shen}, Y., {Grier}, C.~J., {Horne}, K., {et~al.} 2024, \bibinfo{title}{{The Sloan Digital Sky Survey Reverberation Mapping Project: Key Results},} \apjs, 272, 26, \dodoi{10.3847/1538-4365/ad3936}

\bibitem[{P.~S. {Smith} {et~al.}(2009){Smith}, {Montiel}, {Rightley}, {Turner}, {Schmidt}, \& {Jannuzi}}]{Smith2009}
{Smith}, P.~S., {Montiel}, E., {Rightley}, S., {et~al.} 2009, \bibinfo{title}{{Coordinated Fermi/Optical Monitoring of Blazars and the Great 2009 September Gamma-ray Flare of 3C 454.3},} arXiv e-prints, arXiv:0912.3621.
\newblock \doarXiv{0912.3621}

\bibitem[{H. {Ter{\"a}sranta} {et~al.}(1998){Ter{\"a}sranta}, {Tornikoski}, {Mujunen}, {Karlamaa}, {Valtonen}, {Henkel}, {Urpo}, {Lainela}, {Pursimo}, {Nilsson}, {Wireen}, {Roulet}, {Paredes}, {Lara}, \& {Valtaoja}}]{Terasranta1998}
{Ter{\"a}sranta}, H., {Tornikoski}, M., {Mujunen}, A., {et~al.} 1998, \bibinfo{title}{{Fifteen years of monitoring extragalactic radio sources at 22, 37 and 87 GHz},} Astronomy and Astrophysics Supplement Series, 132, 305, \dodoi{10.1051/aas:1998294}

\bibitem[{V. {U} {et~al.}(2022){U}, {Barth}, {Vogler}, {Guo}, {Treu}, {Bennert}, {Canalizo}, {Filippenko}, {Gates}, {Hamann}, {Joner}, {Malkan}, {Pancoast}, {Williams}, {Woo}, {Abolfathi}, {Abramson}, {Armen}, {Bae}, {Bohn}, {Boizelle}, {Bostroem}, {Brandel}, {Brink}, {Channa}, {Cooper}, {Cosens}, {Donohue}, {Fillingham}, {Gonz{\'a}lez-Buitrago}, {Halevi}, {Halle}, {Hood}, {Horne}, {Horst}, {de Kouchkovsky}, {Kuhn}, {Kumar}, {Leonard}, {Loveland}, {Manzano-King}, {McHardy}, {Michel}, {Olaes}, {Park}, {Park}, {Pei}, {Ross}, {Runco}, {Samuel}, {S{\'a}nchez}, {Scott}, {Sexton}, {Shin}, {Shivvers}, {Spencer}, {Stahl}, {Stegman}, {Stomberg}, {Valenti}, {Villafa{\~n}a}, {Walsh}, {Yuk}, \& {Zheng}}]{U2022}
{U}, V., {Barth}, A.~J., {Vogler}, H.~A., {et~al.} 2022, \bibinfo{title}{{The Lick AGN Monitoring Project 2016: Velocity-resolved H{\ensuremath{\beta}} Lags in Luminous Seyfert Galaxies},} \apj, 925, 52, \dodoi{10.3847/1538-4357/ac3d26}

\bibitem[{J.-M. {Wang} {et~al.}(2014){Wang}, {Qiu}, {Du}, \& {Ho}}]{Wang2014}
{Wang}, J.-M., {Qiu}, J., {Du}, P., \& {Ho}, L.~C. 2014, \bibinfo{title}{{Self-shadowing Effects of Slim Accretion Disks in Active Galactic Nuclei: The Diverse Appearance of the Broad-line Region},} \apj, 797, 65, \dodoi{10.1088/0004-637X/797/1/65}

\bibitem[{J.~F.~C. {Wardle} \& P.~P. {Kronberg}(1974){Wardle} \& {Kronberg}}]{wardle1974}
{Wardle}, J.~F.~C., \& {Kronberg}, P.~P. 1974, \bibinfo{title}{{The linear polarization of quasi-stellar radio sources.},} \apj, 194, 249, \dodoi{10.1086/153240}

\bibitem[{J.-H. {Woo} {et~al.}(2024){Woo}, {Wang}, {Rakshit}, {Cho}, {Son}, {Bennert}, {Gallo}, {Hodges-Kluck}, {Treu}, {Barth}, {Cho}, {Foord}, {Geum}, {Guo}, {Jadhav}, {Jeon}, {Kabasares}, {Kang}, {Kim}, {Kim}, {Kim}, {Le}, {Malkan}, {Mandal}, {Park}, {Spencer}, {Shin}, {Sung}, {U}, {Williams}, \& {Yee}}]{Woo2024}
{Woo}, J.-H., {Wang}, S., {Rakshit}, S., {et~al.} 2024, \bibinfo{title}{{The Seoul National University AGN Monitoring Project. III. H{\ensuremath{\beta}} Lag Measurements of 32 Luminous Active Galactic Nuclei and the High-luminosity End of the Size{\textendash}Luminosity Relation},} \apj, 962, 67, \dodoi{10.3847/1538-4357/ad132f}

\bibitem[{C.~C. {Wu}(1977){Wu}}]{Wu1977}
{Wu}, C.~C. 1977, \bibinfo{title}{{Ultraviolet observations of 3C 273 by the ANS.},} \apjl, 217, L117, \dodoi{10.1086/182552}

\bibitem[{Q. {Yuan} {et~al.}(2023){Yuan}, {Kushwaha}, {Gupta}, {Tripathi}, {Wiita}, {Zhang}, {Liu}, {L{\"a}hteenm{\"a}ki}, {Tornikoski}, {Tammi}, {Ramakrishnan}, {Cui}, {Wang}, {Gu}, {Bambi}, \& {Volvach}}]{Yuan2023}
{Yuan}, Q., {Kushwaha}, P., {Gupta}, A.~C., {et~al.} 2023, \bibinfo{title}{{Multiwavelength Temporal Variability of the Blazar PKS 1510-089},} \apj, 953, 47, \dodoi{10.3847/1538-4357/acdd74}

\bibitem[{Z.-X. {Zhang} {et~al.}(2019){Zhang}, {Du}, {Smith}, {et~al.}}]{Zhang2019}
{Zhang}, Z.-X., {Du}, P., {Smith}, P.~S., {et~al.} 2019, \bibinfo{title}{{Kinematics of the Broad-line Region of 3C 273 from a 10 yr Reverberation Mapping Campaign},} \apj, 876, 49, \dodoi{10.3847/1538-4357/ab1099}

\end{thebibliography}

\end{document}